\documentclass[12pt]{article}

\usepackage{epsfig}
\renewcommand{\baselinestretch}{1}
\begin{document}

\title{Wigner Functions with Boundaries}
\author{
Nuno Costa Dias\footnote{{\it nuno.dias@ulusofona.pt}} \\
Jo\~ao Nuno Prata\footnote{{\it joao.prata@ulusofona.pt}} \\  {\it
Departamento de Matem\'{a}tica} \\
{\it
Universidade Lus\'ofona de Humanidades e Tecnologias} \\  {\it Av. Campo Grande, 376,
1749-024 Lisboa, Portugal}}
\maketitle

\begin{abstract}
We consider the general Wigner function for a particle confined to a finite interval and subject to Dirichlet boundary conditions. We derive the boundary corrections to the "star-genvalue" equation and to the time evolution equation. These corrections can be cast in the form of a boundary potential contributing to the total Hamiltonian which together with a subsidiary boundary condition is responsible for the discretization of the energy levels. We show that a completely analogous formulation (in terms of boundary potentials) is also possible in standard operator quantum mechanics and that the Wigner and the operator formulations are also in one-to-one correspondence in the confined case. In particular, we extend Baker's converse construction to bounded systems. Finally, we elaborate on the applications of the formalism to the subject of Wigner trajectories, namely in the context of collision processes and quantum systems displaying chaotic behavior in the classical limit.
\end{abstract}

\section{Introduction}

The Moyal-Wigner-Weyl quantization constitutes a phase space quantization method alternative to canonical or path integral quantizations, \cite{Weyl}-\cite{Takahashi}. Instead of wave functions and operators one deals with quasi-distribution (Wigner) functions and ordinary c-functions in phase-space. The Weyl-symbol associated with a general operator $\hat A (\hat x, \hat p)$ is given by, \cite{Weyl,Lee1,nossoj}:
\begin{equation}
A^W (x,p) \equiv \frac{\hbar}{2 \pi} \int_{- \infty}^{+ \infty} d \xi \int_{- \infty}^{+ \infty} d \eta Tr \left\{ \hat A ( \hat x, \hat p) e^{i \xi \hat x + i \eta \hat p} \right\} e^{-i \xi x - i \eta p}.
\end{equation}
In this framework the average of $\hat A (\hat x, \hat p ;t)$ is evaluated according to the formula: $
< \psi | \hat A (\hat x, \hat p;t) | \psi > = \int_{- \infty}^{+ \infty} dx \int_{- \infty}^{+ \infty} dp A^W (x,p;0) F^W (x,p;t)$, where $F^W(x,p;t)$ is proportional to the Weyl-symbol of the quantum density matrix, $\hat \rho$, and is known as the Wigner function of the system, \cite{Wigner}. For a pure state, $\hat \rho = | \psi>< \psi|$, the Wigner function can be shown to be of the form:
\begin{equation}
F^W (x,p;t) \equiv \frac{1}{ \pi \hbar} \int_{- \infty}^{+ \infty} dy e^{-2i p y / \hbar} \psi^* (x-y;t) \psi (x+y; t),
\end{equation}
where $\psi (x;t)$ is the solution of Schr\"odinger's equation: $
i \hbar \frac{\partial \psi}{\partial t} (x;t) = \hat H  \psi (x;t).$
The Weyl "transform" establishes a biunivocal correspondence between the quantum algebra $\hat{\cal A}$ of observables with standard operator product $\cdot$ and quantum commutator $ \left[ , \right]$, on the one hand, and the "classical" algebra ${\cal A}$ defined over the classical phase space $T^*M$ with a "star-product" $*$ and a Moyal sine-bracket $\left[ ,\right]_M$, on the other hand. The two latter operations are given by, \cite{Moyal,Dias1}:
\begin{equation}
A^W (x,p) * B^W (x,p) = A^W (x,p)  e^{\frac{i \hbar}{2} {\buildrel { \leftrightarrow}\over {\cal J}}} B^W (x,p)= A^W \left( x, p - \frac{i \hbar}{2} {\buildrel { \rightarrow}\over {\partial}_x} \right) B^W \left( x, p + \frac{i \hbar}{2} {\buildrel { \leftarrow}\over {\partial}_x} \right),
\end{equation}
\begin{equation}
\left[ A^W (x,p) , B^W (x,p) \right]_M = \frac{1}{i \hbar} \left( A^W  * B^W  - B^W  * A^W  \right)= \frac{2}{\hbar} A^W (x,p) \sin \left( \frac{\hbar}{2} {\buildrel { \leftrightarrow}\over {\cal J}} \right) B^W (x,p),
\end{equation}
where $
{\buildrel { \leftrightarrow}\over {\cal J}} \equiv \left( {\buildrel { \leftarrow}\over {\frac{\partial}{\partial x}}} {\buildrel { \rightarrow}\over {\frac{\partial}{\partial p}}} - {\buildrel { \leftarrow}\over {\frac{\partial}{\partial p}}} {\buildrel { \rightarrow}\over {\frac{\partial}{\partial x}}} \right)$.
One realizes immediately that in the limit $\hbar \to 0$, the star-product and the Moyal bracket become the ordinary product of c-numbers and the Poisson bracket, respectively.
It is also easy to check that the dynamics of the Wigner function is governed by the Moyal bracket (4):
\begin{equation}
\frac{\partial F^W}{\partial t} (x,p;t) = \left[ H^W (x,p) , F^W (x,p;t) \right]_M,
\end{equation}
where $H^W (x,p)$ is the Weyl symbol of the quantum Hamiltonian, $\hat H ( \hat x, \hat p )$.

If the time-independent wave function happens to be an energy eigenstate with eigenvalue $E$, i.e. $\hat H \psi (x) = E \psi (x)$, then \cite{Fairlie1,Fairlie2,zachos,nossosf} the Wigner function $F^W_E (x,p)$ satisfies an equivalent $*$-genvalue equation with identical $*$-genvalue:
\begin{equation}
H^W (x,p) * F^W_E (x,p) = F^W_E (x,p) * H^W (x,p) = E F^W_E (x,p).
\end{equation}
More generally, to every eigenstate $|a>$ with eigenvalue $a$ of some operator $\hat A$, there is one and only one associated stargenfunction $F^W_a (x,p)$, which is a solution of the stargenvalue equation: $A^W (x,p) * F^W_a (x,p) = a F^W_a (x,p)$ \cite{Baker,Fairlie1,nossosf}. This stargenfunction allows for the evaluation of the probability of measuring the eigenvalue $a$:
\begin{equation}
{\cal P} (A=a) = \int dx \int dp \hspace{0.2 cm} F^W (x,p) F^W_a (x,p).
\end{equation}
It is important to emphasize that, in most cases, the advantage in the Wigner approach does not reside in solving specific problems. Rather, it should be regarded as a means to grasp certain conceptual aspects of quantum mechanics and its connection with classical mechanics. Nevertheless, in some situations it could be a better starting point for finding solutions to specific problems. The Wigner formulation is a useful tool to derive kinetic equations in particular regimes (dilute gas, weakly interacting particles). In nonequilibrium statistical mechanics this is a familiar approach, \cite{Balescu}. In collision processes the Wigner methods are also a useful tool \cite{Lee1,Carruthers}. Contrary to the standard operator approach they display the important advantage of making possible the use of approximation methods similar to the ones used in the full classical treatment. Furthermore the related subject of Wigner trajectories \cite{Lee1,Lee2,Lee3,Lee4} provides a pictorial and practical tool to study collision process and  it is of similar importance in the context of quantum chaotic systems \cite{Lee1,Takahashi,Latka,Shin,Lin}.

Boundary value problems, on the other hand, are one of the cornerstones of quantum mechanics.
These are much more realistic models and moreover they entail a discretization of observables' spectra, which is one of the main features of quantum mechanics. They appear in virtually all branches of physics, ranging from quantum mechanics to general relativity, open strings and D-branes, where the non-commutative *-product and the Moyal bracket also find several applications \cite{Fairlie2,witten,pinzul}.
In standard operator quantum mechanics some famous examples of boundary value problems
include the Kondo problem \cite{Affleck1}, quantum Hall liquids with constriction \cite{Moon} and the Callan-Rubakov model \cite{Affleck2}. Furthermore one of the simplest collision process is the collision against an impenetrable boundary and some standard examples of quantum chaotic systems are bounded \cite{Lee1,Lin}.

In spite of the importance of boundary value problems in general and of these models in particular, Wigner quantum mechanics does not provide a self contained and consistent formulation of confined systems. The aim of this paper is thus to provide such formulation.
More precisely our aim is to generalize the two key equations of Wigner quantum mechanics (eq.(5) and eq.(6)) to the case when the original wave function
is confined to an interval and satisfies Dirichlet boundary conditions.

Let us then consider a general confined system. In standard operator quantum mechanics, we first solve the Schr\"odinger equation in the bulk (with Hamiltonian $\hat H$), subsequently impose the boundary conditions that pin down the physical solutions and finally impose the wave function to be zero beyond the bulk. Alternatively, we may be able to provide a boundary potential $\hat V_D$ that confines the wave-function and then solve the unconstrained system, with extended Hamiltonian $\hat H_D = \hat H + \hat V_D$ and the original conditions at the boundaries.

To obtain the solution of the problem in terms of Wigner functions, we just have to insert the solution obtained in the Schr\"odinger formulation into the definition of the Wigner function. This was the method followed in ref.\cite{Lee2}. But the question that remains is: how do we solve the problem within the Wigner-Weyl formulation?

The first approach cannot be easily translated into the Wigner framework because the highly nonlocal character of the Wigner function produces nontrivial effects of the boundary on the bulk part of the Wigner function, so that the stargenvalue equation has to be altered, as we shall see (sections 2-4).

The second procedure seems to be the natural starting point to derive, through the Weyl map, the Wigner formulation for the bounded system. However this is not an easy task, mainly because the formulation of confined systems in terms of boundary potentials is already problematic at the standard operator level. In section 6 we will consider this approach and a) consistently solve the problem at the standard operator level and b) use the Weyl map to recover the bounded stargenvalue equation derived in section 3.

This paper is organized as follows. In section 2, we discuss some aspects of the boundary conditions satisfied by the Wigner function. In sections 3,4 and 5 we evaluate the boundary corrections to the $*$-genvalue equation (6) and the time evolution equation (5) directly from the general formula of the confined wave function. Furthermore, we a) extend Baker's converse construction to bounded systems and b) argue that the results obtained are compatible with the preservation of the Wigner function's normalization under time evolution. In section 6 we develop the boundary potential approach for standard operator quantum mechanics and show that through the Weyl map this formulation yields the proper Wigner formulation of confined systems. In section 7 we discuss some practical applications of the formalism: the computation of Wigner trajectories for confined chaotic systems and for collision processes. Finally, in section 8 we present our conclusions.

\section{Wigner function in a finite interval}

Let us assume that the domain of the wave function is the finite interval $a < x < b$ with $a<b$, so that the wave function vanishes outside this interval. As we now argue, quantum confined systems cannot be treated, in the Wigner formalism, through the standard procedure of solving a differential equation and imposing the boundary conditions thereafter. In particular, the energy stargenfunctions for these systems do not satisfy the stargenvalue equation (6) with the Hamiltonian of the bulk.

As an example, consider the free particle in the infinite well, with $a=-b = - L/2$. The stargenvalue equation would be of the form:
$$
\frac{p^2}{2m} * F^W (x,p) = E F^W (x,p).
$$
If we solve the corresponding Schr\"odinger equation with Dirichlet boundary conditions we get for the fundamental state:
$$
-\frac{\hbar^2}{2m} \frac{\partial ^2}{\partial x^2} \psi(x) =E\psi(x)  \Longrightarrow \psi_1(x)=\sqrt{\frac{2}{L}} \cos \left(\frac{\pi}{L} x \right).
$$
If we now determine the associated Wigner function using (2) we get \cite{Lee2} that $F^W(x,p)$ is given by eq.(39,40). We can easily check that $F^W(x,p)$ does not satisfy the previous stargenvalue equation and thus the only possible conclusion is that this equation has to be altered. Similarly, we can check that the time-evolution of the system does not satisfy the Moyal equation with bulk Hamiltonian.

Let us start by studying the general properties of the Wigner function for a confined system. Since the wave function vanishes outside the interval $\left]a,b \right[$, it satisfies the condition:
$$
\psi^* (x-y) \psi (x+y) =0 \quad \mbox{unless} \quad a<x-y<b \quad \mbox{and} \quad a<x+y<b.
$$
This means that: 1) for $a<x\le x_0=\frac{a+b}{2}$, we have:
$$
\psi^* (x-y) \psi (x+y) =0 \quad \mbox{unless} \quad a-x<y<x-a ,
$$
similary, 2) for $x_0 < x<b$:
$$
\psi^* (x-y) \psi (x+y) =0 \quad \mbox{unless} \quad x-b<y<b-x,
$$
and finally, 3) for $x \le a$ or $x\ge b$:
$$
\psi^* (x-y) \psi (x+y) =0, \quad \forall y.
$$
We conclude that the Wigner function (2) for the bounded system can be written as:
\begin{equation}
F^W (x,p) = \frac{1}{\pi \hbar} \int_{- \infty}^{+ \infty} dy e^{- 2ipy / \hbar} \psi^* (x-y) \psi (x+y) = \Theta_1 (x) F^W_1 (x,p) + \Theta_2 (x) F^W_2 (x,p),
\end{equation}
where:
\begin{equation}
\begin{array}{l}
\Theta_1 (x) = \theta (x-a) - \theta (x -x_0)  = \left\{
\begin{array}{l l}
1, & \mbox{if } a < x \le x_0 \\
& \\
0, & \mbox{otherwise}
\end{array}
\right.\\
\\
\Theta_2 (x) = \theta (x-x_0) + \theta (b - x) -1= \left\{
\begin{array}{l l}
1, & \mbox{if } x_0 < x < b \\
& \\
0, & \mbox{otherwise}
\end{array}
\right.
\end{array},
\end{equation}
$\theta(x)$ is the Heaviside's step function: $\theta(x)=1$ iff $x>0$ and $\theta(x)=0$ otherwise, and finally:
\begin{equation}
\left\{
\begin{array}{l}
F^W_1 (x,p) = \frac{1}{ \pi \hbar} \int_{a-x}^{x - a} dy  e^{-2i p y / \hbar} \psi^* (x-y) \psi (x+y),\\
\\
F^W_2 (x,p) =  \frac{1}{ \pi \hbar} \int_{x-b}^{b-x} dy  e^{-2i p y / \hbar} \psi^* (x-y) \psi (x+y)
\end{array}
\right..
\end{equation}
Since the wave function or its derivatives are possibly discontinuous at $x=a$ and $x=b$, and to avoid possible misinterpretations in our future calculations, the previous integrals are defined as improper: $\int_{a-x}^{x - a}$ stands for $\lim_{c \to a^+} \int_{c-x}^{x - c}$ and likewise $\int_{x-b}^{b-x}$ stands for $\lim_{c\to b^-} \int_{x-c}^{c-x}$. Notice that this is fully compatible with our previous results identifying the domain where $\psi^*(x-y)\psi(x+y)$ is not identically zero.

As a second remark let us point out that the boundary conditions on $\psi(x)$ are imposed on the bulk side of the boundary. For instance, the Dirichlet boundary conditions would be of the form: $\lim_{\epsilon \to 0^+} \psi(a+\epsilon) = \lim_{\epsilon \to 0^+} \psi(b-\epsilon) = 0$, which is not the same as requiring $\psi(a)=\psi(b)=0$, since these are valid independently of the boundary conditions satisfied by $\psi(x)$. This is so because $\psi(x)$ is confined to the open interval $]a,b[$. It is important to realize that we could equally choose to defined "confined" as "confined to the close interval $[a,b]$" and in fact for Dirichlet boundary conditions the two prescriptions yield exactly the same wave function.
However, the confinement to an open interval makes some of the future steps more natural (though occasionally more involved) and this is why we choose to work with this prescription.

Finally, and to make our future expressions more compact, we introduce the notation: $f(c^{\pm})=\lim_{\epsilon \to 0^+} f(c\pm \epsilon)$ which will be used whenever there is no risk of misunderstanding.

Let us then study the boundary conditions that are satisfied by the Wigner function of a confined system. From (8-10), one realizes immediately that:
\begin{equation}
\left\{
\begin{array}{l}
F^W(a^+,p)= F^W_1 (a^+,p) = F^W_2 (b^-,p)= F^W(b^-,p)= 0,\\
 \\
F^W (x_0^-,p)= F^W_1 (x_0,p) = F^W_2 (x_0,p) =F^W (x_0^+,p),
\end{array}
\right.
\end{equation}
where, accordingly to the previous notation, $F^W(a^+,p)$ stands for the limit $\lim_{\epsilon \to 0^+} F^W(a+\epsilon,p)$ and likewise for the other expressions. Since $F^W(x,p)$ is defined as an improper integral it already requires the evaluation of a limit. This limit will always be calculate before the limit $\lim_{x \to a^+} F^W(x,p)$, i.e. first we evaluate the Wigner function for all $x$ and only then do we compute whatever limits of the Wigner function.
Let us proceed: eq.(11) means that $F^W (x,p)$ is continuous at $x_0$ and obeys Dirichlet boundary conditions irrespective of the particular boundary conditions satisfied by the associated wave function $\psi$. The boundary conditions obeyed by the confined wave function act only on the derivatives of the Wigner function. For $a<x \le x_0$ a straightforward computation leads to:
\begin{equation}
\begin{array}{c}
 \frac{\partial F^W_1 }{\partial x} (x,p) = \frac{1}{\pi \hbar} \left\{ e^{-\frac{2ip}{\hbar} (x-a)} \psi^* (a^+) \psi (2x-a^+) +  e^{-\frac{2ip}{\hbar} (a-x)} \psi^* (2x -a^+) \psi (a^+) + \right.\\
\\
\left. + \int_{a-x}^{x-a} dy e^{-2i py / \hbar} \left[ \psi'^* (x-y) \psi (x+y) + \psi^* (x-y) \psi' (x+y) \right] \right\},
\end{array}
\end{equation}
where the prime denotes the derivative with respect to the argument. Consequently:
\begin{equation}
\begin{array}{l l}
\frac{\partial F^W }{\partial x} (a^+,p) & = \lim_{x \to a^+} \frac{\partial F_1^W }{\partial x} (x,p) =\frac{2}{\pi \hbar} | \psi (a^+)|^2\\
& \\
\frac{\partial F^W }{\partial x} (x_0^- ,p) & = \frac{\partial F_1^W }{\partial x} (x_0 ,p)=
\frac{1}{\pi \hbar} \left[ e^{-\frac{ip}{\hbar} (b-a)} \psi^* (a^+) \psi (b^-) +  e^{-\frac{ip}{\hbar} (a-b)} \psi^* (b^-) \psi (a^+)\right] +\\
& \\
& + \frac{1}{\pi \hbar} \lim_{\epsilon \to 0^+} \int_{\frac{a-b}{2}+\epsilon}^{\frac{b-a}{2}-\epsilon} dy e^{-2i py / \hbar} \left[ \psi'^* (x_0-y) \psi (x_0+y) + \psi^* (x_0-y) \psi' (x_0+y) \right].
\end{array}
\end{equation}
Similarly, from $F_2^W$ we get:
\begin{equation}
\begin{array}{l l}
\frac{\partial F^W }{\partial x} (b^-,p) & = \lim_{x \to b^-} \frac{\partial F_2^W }{\partial x} (x,p) =- \frac{2}{\pi \hbar} | \psi (b^-)|^2\\
& \\
\frac{\partial F^W }{\partial x} (x_0^+ ,p) & = \frac{\partial F_2^W }{\partial x} (x_0 ,p)=
- \frac{1}{\pi \hbar} \left[ e^{-\frac{ip}{\hbar} (b-a)} \psi^* (a^+) \psi (b^-) +  e^{-\frac{ip}{\hbar} (a-b)} \psi^* (b^-) \psi (a^+)\right] +\\
& \\
& + \frac{1}{\pi \hbar}\lim_{\epsilon \to 0^+} \int_{\frac{a-b}{2}+\epsilon}^{\frac{b-a}{2}-\epsilon} dy e^{-2i py / \hbar} \left[ \psi'^* (x_0-y) \psi (x_0+y) + \psi^* (x_0-y) \psi' (x_0+y) \right].
\end{array}
\end{equation}
We conclude that in general $ \frac{\partial F^W }{\partial x} (x_0^- ,p) \ne \frac{\partial F^W }{\partial x} (x_0^+ ,p)$, except for Dirichlet boundary conditions (i.e. $\psi (a^+) = \psi^*  (a^+) = \psi (b^-) = \psi^*  (b^-)=0$), in which case the Wigner function also obeys Neumann boundary conditions: $\frac{\partial F^W }{\partial x} (a^+ ,p) = \frac{\partial F^W }{\partial x} (b^- ,p)=0$. We can carry on with this process and compute the second and third derivatives. The results concerning the boundary conditions can then be summarized as follows. All Wigner functions for bounded systems satisfy eq.(11). These are consistency conditions and, given their general character, cannot give countenance to the discretization of the energy spectrum. The boundary conditions satisfied by the confined wave function yield subsidiary conditions on the Wigner function:

\vspace{0.3 cm}
\noindent
{\bf Dirichlet}: $\psi (a^+) = \psi^*  (a^+) = \psi (b^-) = \psi^*  (b^-)=0$
\begin{equation}
\frac{\partial F^W }{\partial x} (a^+,p) = \frac{\partial F^W }{\partial x} (b^-,p) = 0.
\end{equation}
If a Wigner function satisfies these Neumann conditions, then it {\it automatically} verifies:
\begin{equation}
\left\{
\begin{array}{l}
\frac{\partial F^W }{\partial x} (x_0^-,p) =  \frac{\partial F^W }{\partial x} (x_0^+,p),\\
\\
\frac{\partial^2 F^W }{\partial x^2} (a^+,p) = \frac{\partial^2 F^W }{\partial x^2} (b^-,p)=0,\\
 \\
\frac{\partial^2 F^W }{\partial x^2} (x_0^-,p) =  \frac{\partial^2 F^W }{\partial x^2} (x_0^+,p).
\end{array}
\right.
\end{equation}
Obviously, this pattern does not continue indefinitely. For instance, $\frac{\partial^3 F^W }{\partial x^3} (a^+,p) = \frac{8}{\pi \hbar} |\psi' (a^+)|^2 \ne 0$. It is also important to emphasize that contrary to eq.(15), the additional set of eqs.(16), being just a consequence of eq.(15), do not constrain $F^W$ any further.

\vspace{0.3 cm}
\noindent
{\bf Neumann}: $\psi' (a^+) = \psi'^*  (a^+) = \psi' (b^-) = \psi'^*  (b^-)=0$
\begin{equation}
\left\{
\begin{array}{l}
\frac{\partial F^W }{\partial x} (a^+,p) \ne 0, \qquad \frac{\partial F^W }{\partial x} (b^-,p) \ne 0\\
\\
\frac{\partial F^W }{\partial x} (x_0^-,p) \ne  \frac{\partial F^W }{\partial x} (x_0^+,p)\\
 \\
\frac{\partial^2 F^W }{\partial x^2} (a^+,p) = \frac{\partial^2 F^W }{\partial x^2} (b^-,p)=0\\
\\
\frac{\partial^2 F^W }{\partial x^2} (x_0^-,p) =  \frac{\partial^2 F^W }{\partial x^2} (x_0^+,p).
\end{array}
\right.
\end{equation}
We also have: $\frac{\partial^3 F^W }{\partial x^3} (a^+,p ) \ne 0$.
\\

The former results show that the confinement of the system and the boundary conditions satisfied by the wave function influence the Wigner function in a very non-trivial way. Furthermore, the conditions (11,15,16) will play an important part in the derivation of the new $*$-genvalue equation for bounded systems.

However, one should realize that the previous results also display some less interesting features: let us focus on the case in which the wave function satisfies Dirichlet boundary conditions. If the system is also confined then the Wigner function will satisfy the set of conditions (11,15,16). The problem is that the converse result is not valid, i.e. the conditions (11,15,16) do not imply the confinement of the system and it is also not clear if, in general, they imply that the original wave function satisfy Dirichlet boundary conditions. Hence, (11,15,16) will not completely pin down the Wigner transform of the confined wave function satisfying Dirichlet boundary conditions and thus eqs.(11,15,16) fail to provide the proper translation (into the Wigner language) of the Dirichlet boundary conditions on the wave function.

We now introduce a new set of boundary conditions for the Wigner function, which are fully equivalent to Dirichlet boundary conditions on the original wave function, both for the confined as well as for the unconfined case.
We call these conditions {\it integral Dirichlet boundary conditions} and they can be derived as follows: if the wave function satisfies Dirichlet boundary conditions then $\lim_{\epsilon \to 0+} {\cal P}(a+\epsilon) = \lim_{\epsilon \to 0+} |\psi(a+\epsilon)|^2=0$ and similarly $\lim_{\epsilon \to 0+} {\cal P}(b-\epsilon) = \lim_{\epsilon \to 0+} |\psi(b-\epsilon)|^2=0$
where ${\cal P}(x)$ stands for the probability distribution. These equations can be written in the Wigner context quite straightforwardly:
\begin{equation}
\left\{
\begin{array}{l}
\lim_{\epsilon \to 0+} \int_{-\infty}^{+\infty} dp F^W(a+\epsilon,p) =0,\\
\\
\lim_{\epsilon \to 0+} \int_{-\infty}^{+\infty} dp F^W(b-\epsilon,p) =0.
\end{array}
\right.
\end{equation}
Notice that the use of the limits is valid (though unnecessary) in the unconfined case (where the wave function and its derivatives are continuous) but it is necessary if the system is confined.

From the two sets of boundary conditions satisfied by the Wigner function, (eqs.(11,15,16) and eqs.(18)), the Dirichlet integral ones are the most appealing both from the physical and the mathematical points of view: firstly, because they represent a physically meaningful imposition, that the probability of finding the particle at the boundary is zero. Secondly, because they are fully equivalent to the Dirichlet boundary conditions on the wave function. From eq.(7,18), we get:
\begin{equation}
\lim_{\epsilon \to 0+} \int_{-\infty}^{+\infty} dp F^W(a+\epsilon,p) =0 \Longrightarrow
\lim_{\epsilon \to 0+} |\psi(a+\epsilon)|^2 =0 \Longrightarrow \lim_{\epsilon \to 0+} \psi(a+\epsilon)=0,
\end{equation}
and similarly for the $b$-boundary. Finally, and to avoid a possible misinterpretation let us point out that eq.(11) does not imply
the satisfaction of the conditions (18) given the fact that, in general, it is not possible to interchange the order in which one evaluates the limit and the integral in (18). For instance, if a confined wave function satisfies Neumann conditions then the Wigner function will satisfy (11) but not (18).

\section{Boundary stargenvalue equation}

Let us henceforth assume that the wave function satisfies Dirichlet boundary conditions, so that eqs.(11,15,16,18) are valid. Furthermore, we assume that $\psi (x)$ is an energy-eigenstate for a Hamiltonian of the form: $\hat H (\hat x, \hat p)= \frac{\hat p^2}{2m} + V(\hat x)$. If we substitute $F^W$ (8) on the left-hand side of equation (6), we get:
\begin{equation}
H^W  * F^W= H^W * \left\{\Theta_1 F_1^W + \Theta_2 F_2^W \right\}=
\left\{\frac{p^2}{2m} + V(x) \right\} * \left( \Theta_1 F_1^W \right)
+\left\{\frac{p^2}{2m} + V(x) \right\} * \left( \Theta_2 F_2^W \right).
\end{equation}
A straightforward calculation leads to:
\begin{equation}
\begin{array}{c}
H^W * \left( \Theta_1 F^W_1 \right) =  \Theta_1 \left(H^W * F^W_1 \right) - \frac{i \hbar p}{2m} \left[ \delta (x-a) - \delta (x - x_0) \right] F^W_1 +\\
\\
+ \frac{\hbar^2}{8m} \left[ \delta' (x - x_0) - \delta' (x-a) \right] F^W_1 + \frac{\hbar^2}{4m} \left[ \delta (x - x_0) - \delta (x-a) \right] \frac{\partial F^W_1}{\partial x},
\end{array}
\end{equation}
taking into account the boundary conditions at $x=a$, we get:
\begin{equation}
\begin{array}{c}
H^W * \left( \Theta_1 F^W_1 \right) =  \Theta_1 \left(H^W* F^W_1 \right) + \frac{i \hbar p}{2m}  \delta (x - x_0)  F^W_1 (x_0 , p) + \\
\\
+\frac{\hbar^2}{4m} \delta (x - x_0) \frac{\partial F^W_1}{\partial x} (x_0,p) +  \frac{\hbar^2}{8m}  \delta' (x - x_0)  F^W_1 (x,p) .
\end{array}
\end{equation}
Similarly, we also have:
\begin{equation}
\begin{array}{c}
H^W * \left( \Theta_2 F^W_2 \right) = \Theta_2 \left(H^W* F^W_2 \right)   - \frac{i \hbar p}{2m}  \delta (x - x_0)  F^W_2 (x_0 , p) -\\
\\
- \frac{\hbar^2}{4m} \delta (x - x_0) \frac{\partial F^W_2}{\partial x} (x_0,p) -  \frac{\hbar^2}{8m}  \delta' (x - x_0)  F^W_2 (x,p) .
\end{array}
\end{equation}
Adding up all contributions and taking into account the fusing conditions (11) and (16), we obtain:
\begin{equation}
H^W  * F^W= \Theta_1 \left(H^W* F_1^W \right)+ \Theta_2 \left( H^W* F_2^W \right).
\end{equation}
Let us then calculate the term $H^W* F_1^W $ for $a<x\le x_0$ and the term $H^W* F_2^W $ for $x_0<x<b$. The following theorem will do this.\\

\underline{\bf Theorem}\\
Let $F^W_1$ and $F_2^W$ be given by eq.(10) where $\psi(x)$ is an energy eigenstate satisfying Dirichlet boundary conditions at $a$ and $b$. Let $H^W=\frac{p^2}{2m} +V(x)$. We then have:
\begin{equation}
\left\{ \begin{array}{l}
H^W * F_1^W (x,p) = E F_1^W (x,p)+ \frac{\hbar^2}{2m} \delta' (x-a) * F_1^W (x^+,p), \quad \mbox{for} \quad a<x\le x_0  \\
\\
H^W * F_2^W (x,p) = E F_2^W (x,p)-\frac{\hbar^2}{2m} \delta' (x-b) * F_2^W (x^-,p), \quad \mbox{for} \quad x_0<x< b
\end{array} \right.
\end{equation}

\underline{\bf Proof}\\
Let us consider the product $H^W * F_1^W$. Following the method described in \cite{Fairlie1}, we get:
\begin{equation}
\begin{array}{c}
H^W  * F^W_1  = \frac{1}{\pi \hbar} \left[ \frac{1}{2m} \left( p - \frac{i \hbar}{2}
{\buildrel { \rightarrow}\over {\partial}_x} \right)^2 + V(x) \right] \int_{a-x}^{x-a} dy e^{- \frac{2i y}{\hbar} \left( p + \frac{i \hbar}{2} {\buildrel {\leftarrow}\over{\partial}_x} \right)} \psi^* (x-y) \psi (x+y) =\\
\\
= \frac{1}{\pi \hbar}  \int_{a-x}^{x-a} dy
e^{- \frac{2i yp}{\hbar} } V(x+y) \psi^* (x-y) \psi (x+y) +\\
\\
+\frac{1}{2m \pi \hbar} \left[ p^2 - i \hbar p {\buildrel { \rightarrow}\over {\partial}_x} - \frac{\hbar^2}{4} {\buildrel { \rightarrow}\over {\partial}_x}^2 \right]
\int_{a-x}^{x-a} dy  e^{- \frac{2i yp}{\hbar} }  \psi^* (x-y) \psi (x+y).
\end{array}
\end{equation}
where ${\buildrel { \rightarrow}\over {\partial}_x}$ acts on $F_1^W$ and
${\buildrel { \leftarrow}\over {\partial}_x}$ acts on $H^W$.
Taking into account that $p e^{- \frac{2i yp}{\hbar} } = \frac{i \hbar}{2} \frac{\partial}{\partial y} e^{- \frac{2i yp}{\hbar} }$ and that $\psi (x)$ is an energy eigenstate, we get after a few integrations by parts:
\begin{equation}
H^W (x,p) * F^W_1 (x,p) = E F^W_1 (x,p) + {\cal B}_1 (x,p).
\end{equation}
The extra term ${\cal B}_1 (x,p)$ is a boundary correction given by:
\begin{equation}
{\cal B}_1 (x,p) = - \frac{\hbar}{2 \pi m} e^{-\frac{2ip}{\hbar} (a-x)} \left\{\frac{2ip}{\hbar}  \psi^* (2x-a^-) \psi (a^+) +  \psi'^* (2x-a^-) \psi (a^+) + \psi^* (2x-a^-) \psi' (a^+) \right\}.
\end{equation}
For Dirichlet boundary conditions, we get:
\begin{equation}
{\cal B}_1^D (x,p) = - \frac{\hbar}{2 \pi m} e^{-\frac{2ip}{\hbar} (a-x)}  \psi^* (2x-a^-) \psi' (a^+) .
\end{equation}

Let us now attempt to express ${\cal B}_1^D (x,p)$ in terms of the Wigner function $F^W_1 (x,p)$. Consider the following integral (where $\epsilon >0$):
\begin{equation}
\begin{array}{c}
\Lambda_{\epsilon} \equiv \int_{ - \infty}^{+ \infty} dk  e^{ik (x-a)} k  F^W_1 \left(x+{\epsilon}, p - \frac{\hbar}{2} k \right)= \\
\\
=\frac{1}{\pi \hbar} \int_{ - \infty}^{+ \infty} dk \int_{a-x-{\epsilon}}^{x-a+{\epsilon}} dy e^{ik (x-a+y)} k  e^{-2ipy/ \hbar} \psi^* (x+{\epsilon}-y) \psi (x+{\epsilon}+y) =\\
\\
=\frac{2}{i \hbar} \int_{a-x-{\epsilon}}^{x-a+{\epsilon}}  dy  \left[ \frac{\partial }{\partial y} \delta (x-a+y) \right]   e^{-2ipy/ \hbar} \psi^* (x+{\epsilon}-y) \psi (x+{\epsilon}+y) =  \\
\\
=-\frac{2}{i \hbar} \int_{a-x-{\epsilon}}^{x-a+{\epsilon}}  dy  \delta (x-a+y) e^{-2ipy/ \hbar}
\left\{ - \frac{2ip}{\hbar} \psi^*(x+{\epsilon}-y) \psi (x+{\epsilon}+y) - \right.\\
\\
\left. \psi'^*(x+{\epsilon}-y) \psi (x+{\epsilon}+y)  +  \psi^*(x+{\epsilon}-y) \psi' (x+{\epsilon}+y) \right\}= \\
\\
= \frac{2i}{\hbar} e^{- \frac{2ip}{\hbar} (a-x)} \left\{ - \frac{2ip}{\hbar} \psi^*(2x-a+{\epsilon}) \psi (a+{\epsilon})  - \psi'^*(2x-a+{\epsilon}) \psi (a+{\epsilon})  +  \psi^*(2x-a+{\epsilon}) \psi' (a+{\epsilon})\right\}.
\end{array}
\end{equation}
In particular for a wave function satisfying a Dirichlet boundary condition, we conclude that: $
\lim_{{\epsilon} \to 0^+} \Lambda^D_{\epsilon} (x,p;a) = \frac{2i}{\hbar} e^{-\frac{2ip}{\hbar} (a-x)} \psi^*(2x-a^-) \psi'(a^+)$.
It is straightforward to obtain from eq.(29): $
{\cal B}_1^D (x,p) = \lim_{{\epsilon} \to 0^+} \frac{ i \hbar^2}{4 \pi m} \Lambda^D_{\epsilon} (x,p;a )$. Let us now try to cast this expression in terms of $*$-products:
\begin{equation}
\begin{array}{c}
\Lambda_{\epsilon} (x,p) = \sum_{n=0}^{\infty} \frac{1}{n!} \left( - \frac{\hbar}{2} \right)^n \frac{ \partial^n F^W_1 (x+{\epsilon},p)}{\partial p^n}  \int_{- \infty}^{+ \infty} dk e^{ik (x-a)} k^{n+1}= \\
\\
= -2 i \pi \sum_{n=0}^{\infty} \frac{1}{n!} \left( \frac{i \hbar}{2} \right)^n \delta^{(n+1)} (x-a) \frac{\partial^n F^W_1}{\partial p^n} (x+{\epsilon},p) = -2 i \pi \left[ \delta' (x-a) * F^W_1 (x+{\epsilon},p ) \right].
\end{array}
\end{equation}
Taking the limit $\epsilon \to 0^+$ of the previous expression and substituting it in (29) and (27) yields the first equation of (25) directly. As a final remark we notice that $\delta' (x-a) * F^W_1 (x^+,p )=\lim_{\epsilon \to 0} \{\delta' (x-a) * F^W_1 (x+{\epsilon},p ) \}$ is not identical to $\delta' (x-a) * \{ \lim_{\epsilon \to 0} F^W_1 (x+{\epsilon},p ) \} = \delta' (x-a) * F^W_1 (x,p )$. This last expression is not even well defined (cf.(30)) since the Dirac delta function is evaluated at a point where $\psi'(x)$ is not continuous. Hence, we should keep in mind that in general, we cannot interchange the order in which the starproduct and the limit $\epsilon \to 0$ are evaluated.

Finally, we perform the analogous procedure for $F^W_2$ and get the second equation of (25), which concludes the proof.
\\

Let us then return to equation (24). Using eq.(25) we get:
\begin{equation}
\begin{array}{c}
H^W(x,p)  * F^W(x,p)= \Theta_1(x) \left\{E F_1^W(x,p) +\frac{\hbar^2}{2m} \delta'(x-a) * F_1^W(x^+,p) \right\}+ \\
\\
+\Theta_2(x) \left\{ E F_2^W(x,p) - \frac{\hbar^2}{2m} \delta'(x-b) * F_2^W(x^-,p) \right\}.
\end{array}
\end{equation}
Let us try to find the corresponding compact version for $F^W(x,p)$ (eq.(8)).
We notice that:\\
{\bf 1)} $\Theta_1 E F_1^W + \Theta_2 E F_2^W = E \left\{\Theta_1 F_1^W + \Theta_2 F_2^W \right\} =EF^W$.\\
{\bf 2)}
\begin{equation}
\begin{array}{c}
\Theta_1 (x) \left\{ \delta'(x-a) * F_1^W (x^+,p) \right\} = \Theta_1(x) \left\{
- \frac{1}{ \pi \hbar} e^{-\frac{2ip}{\hbar} (a-x)}  \psi^* (2x-a^-) \psi' (a^+)  \right\}=\\
\\
=\Theta_1(x^+) \left\{
- \frac{1}{ \pi \hbar} e^{-\frac{2ip}{\hbar} (a-x)}  \psi^* (2x-a^-) \psi' (a^+)  \right\}=
\delta'(x-a)* \left\{ \Theta_1 (x^+)  F_1^W (x^+,p) \right\},
\end{array}
\end{equation}
where in the first step we used eqs.(29,31); in the second step the fact that $\Theta_1(x) = \lim_{\epsilon \to 0^+} \Theta_1(x+\epsilon), \forall x\not=a,x_0$ and that $\psi^*(2x-a^-)$ vanishes for $x=a$ and $x=x_0$; and finally in the last step that $\Theta_1(x^+) $ is $p$-independent.\\
{\bf 3)} Similarly, we have: $\Theta_2 (x) \left\{ \delta'(x-b) * F_2^W (x^-,p) \right\}=
\delta'(x-b) *\left\{ \Theta_2 (x^-)  F_2^W (x^-,p) \right\}$.\\
{\bf 4)} Finally, we also have:
\begin{equation}
\begin{array}{c}
\delta'(x-b) *\left\{ \Theta_1 (x^+)  F_1^W (x^+,p) \right\}=
\Theta_1 (x^+) \left\{ \delta'(x-b) * F_1^W (x^+,p) \right\}=\\
\\
= \frac{i}{2 \pi} \Theta_1 (x^+) \int_{- \infty}^{+ \infty} dk e^{ik (x-b)} k F^W_1 \left( x^+, p - \frac{\hbar}{2} k \right)
= -  \frac{1}{\pi \hbar} \Theta_1 (x^+) \theta(x^+-x_0) e^{- \frac{2ip}{\hbar} (b-x)} \times \\
\\
\times \left\{ -\frac{2ip}{\hbar} \psi^* (2x -b^-) \psi (b^+) -\psi'^* (2x -b^-) \psi (b^+)+\psi^* (2x -b^-) \psi' (b^+) \right\}=0,
\end{array}
\end{equation}
where in the third step we made the same calculation as in eq.(30) and in the last step we used the fact that $\psi (b^+)=\psi'(b^+)=0 $, or else that $\Theta_1(x^+)=0, \forall x\ge x_0$ and $\theta (x^+-x_0)=0, \forall x<x_0$. \\
{\bf 5)} Similarly:
\begin{equation}
\delta'(x-a) * \left( \Theta_2(x^-) F^W_2 (x^-,p) \right)  =0.
\end{equation}

Using these results we can re-write eq.(32) as:
\begin{equation}
\begin{array}{c}
H^W * F^W= E F^W + \frac{\hbar^2}{2m} \delta'(x-a)* \left\{ \Theta_1 (x^+)  F_1^W (x^+,p) +
\Theta_2 (x^-)  F_2^W (x^-,p) \right\} -\\
\\
-\frac{\hbar^2}{2m} \delta'(x-b)* \left\{ \Theta_1 (x^+)  F_1^W (x^+,p) +
\Theta_2 (x^-)  F_2^W (x^-,p) \right\},
\end{array}
\end{equation}
from where it immediately follows:
\begin{equation}
\left(\frac{p^2}{2m} + V(x) \right) *F^W(x,p) - \frac{\hbar^2}{2m} \delta' (x-a) *F^W(x^+,p)
+\frac{\hbar^2}{2m} \delta' (x-b) *F^W(x^-,p)
= E F^W (x,p).
\end{equation}
Following the same procedure, one would also obtain:
\begin{equation}
F^W(x,p)*\left(\frac{p^2}{2m} + V(x) \right)  - \frac{\hbar^2}{2m} F^W(x^+,p)* \delta' (x-a)
+\frac{\hbar^2}{2m} F^W(x^-,p)*\delta' (x-b)
= E F^W (x,p).
\end{equation}
We conclude that, if the eigenfunction satisfies Dirichlet boundary conditions, then the corresponding Wigner function does not satisfy the standard $*$-genvalue equation (6), but rather a modified equation given by (37,38). Furthermore, the source of discretization of the energy spectrum are not the Dirichlet conditions on the Wigner function but instead the subsidiary boundary conditions (15) or (18).

As a simple example let us consider the free particle in the infinite potential well. We choose as Ansatz:
\begin{equation}
\begin{array}{c}
F^W (x,p) = - \frac{ \alpha^2 }{2 \pi p} \cos \left[ 2 ( \beta - k |x| ) \right] \cdot \sin \left[ \frac{2 p}{ \hbar} \left( \frac{L}{2} - |x| \right) \right] + \\
\\
+ \frac{\alpha^2}{4 \pi (p+ \hbar k) } \sin \left[ \frac{2}{\hbar} \left( p + \hbar k \right) \left( \frac{L}{2} - |x| \right) \right] + \frac{\alpha^2}{4 \pi (p- \hbar k) } \sin \left[ \frac{2}{\hbar} \left( p - \hbar k \right) \left( \frac{L}{2} - |x| \right) \right] ,
\end{array}
\end{equation}
corresponding to the wave-function $\psi (x) = \alpha \sin (k x + \beta)$. This function is continuous at $x_0 = 0$ and satisfies Dirichlet's condition at $x= \pm \frac{L}{2}$, in accordance with (11). If we impose the constraints (15) and substitute this expression in the $*$-genvalue equation (37,38), using the integral form $\Lambda$ (eq.(30)), we obtain:
\begin{equation}
k_n = \frac{n \pi}{L}, \qquad \beta_n = \frac{n \pi}{2}, \qquad E_n = \frac{ \hbar^2 k_n^2}{2m} = \frac{\hbar^2 \pi^2}{2m L^2} n^2, \qquad (n=1,2, \cdots).
\end{equation}
Finally, if we impose the normalization $\int_a^b dx \int_{- \infty}^{+ \infty} dp F^W (x,p) =1$, we obtain $\alpha = \sqrt{2/L}$. This result is in perfect agreement with ref.\cite{Lee2}.

\section{Baker's converse construction}

In the last section we proved that if the wave function is confined to the interval $]a,b[$, satisfies the operator eigenvalue equation and Dirichlet boundary conditions at $a^+$ and $b^-$ then the corresponding Wigner function satisfies a modified *-genvalue equation together with integral Dirichlet boundary conditions at $a^+$ and $b^-$. In this section we want to prove the converse result. In the unconfined case this result is known as Baker's converse construction \cite{Baker,Fairlie1}. Hence, in this section we want to extend Baker's converse construction to the confined case thus proving the full equivalence of the Wigner and operator formulations of quantum mechanics when boundaries are present.

Let us then consider some real function $F^W (x,p)$ satisfying the following conditions:

\noindent
(i) it is a continuous function of $x$ and $p$;

\noindent
(ii) it has an infinite number of derivatives with respect to $p$ and is twice differentiable with respect to $x$;

\noindent
(iii) $F^W$ obeys integral Dirichlet boundary conditions at $x=a^+$ and $x=b^-$, eq.(18);

\noindent
(iv) it obeys the left- and right- boundary stargenvalue equations (37,38) when $\epsilon \to 0^+$:
\begin{equation}
\begin{array}{c}
\left( \frac{p^2}{2m} + V(x) \right) * F^W (x,p) - \frac{\hbar^2}{2m} \delta' (x-a) * F^W(x+ \epsilon, p) +\frac{\hbar^2}{2m} \delta' (x-b) * F^W(x- \epsilon, p)= \\
\\
= F^W (x,p)* \left( \frac{p^2}{2m} + V(x) \right) - \frac{\hbar^2}{2m}  F^W(x+ \epsilon, p)* \delta' (x-a) + \frac{\hbar^2}{2m}  F^W(x- \epsilon, p)* \delta' (x-b)= E F^W (x,p).
\end{array}
\end{equation}
We will now prove that under these assumptions, there is a unique (up to a global phase factor) normalized complex and continuous function $\psi$, related to $F^W$ according to (2), which obeys Dirichlet boundary conditions at $x=a^+$ and $x=b^-$ and is a solution of Schr\"odinger's equation.

It will prove useful to consider the Fourier transform of $F^W$:
\begin{equation}
F^W (x,p) = \frac{1}{\pi \hbar} \int_{- \infty}^{+ \infty} dy \hspace{0.2 cm} e^{- 2ipy / \hbar} \tilde F (x,y).
\end{equation}
Substituting this expression in the left- and right- stargenvalue equations (41), and following the same steps as in eq.(26), we obtain:
\begin{equation}
\begin{array}{c}
\frac{1}{\pi \hbar} \int_{- \infty}^{+ \infty} dy \hspace{0.2 cm} e^{- 2ipy / \hbar} \left\{\left[ - \frac{\hbar^2}{2m} \left( \frac{\partial_x \pm \partial_y}{2} \right)^2 + V (x \pm y) - E \right] \tilde F (x,y) \right. \\
\\
\left. - \frac{\hbar^2}{2m}\delta' (x \pm y-a) \tilde F (x +\epsilon,y)
+ \frac{\hbar^2}{2m}\delta' (x \pm y-b) \tilde F (x -\epsilon,y) \right\}=0.
\end{array}
\end{equation}
This means that $\tilde F (x,y)$ satisfies the equations:
\begin{equation}
\left[ - \frac{\hbar^2}{2m} \left(\frac{\partial_x \pm \partial_y }{2}\right)^2 + V (x\pm y) \right] \tilde F (x,y) - \frac{\hbar^2}{2m} \delta' (x \pm y-a) \tilde F (x +\epsilon,y)
+\frac{\hbar^2}{2m} \delta' (x \pm y-b) \tilde F (x -\epsilon,y)= E \tilde F (x,y).
\end{equation}
From now on we will follow the steps of the standard Baker converse construction for unconfined systems \cite{Baker}. Introducing the function $F$ such that $F(x-y,x+y)=\tilde F(x,y)$ and performing the change of variables $u=x-y$ and $v=x+y$, we get:
\begin{equation}
\left\{ \begin{array}{l}
\left[ - \frac{\hbar^2}{2m} \frac{\partial^2}{\partial v^2} + V (v) \right] F (u,v) - \frac{\hbar^2}{2m} \delta' (v-a)  F (u +\epsilon,v+\epsilon)
+\frac{\hbar^2}{2m} \delta' (v-b) F (u -\epsilon,v-\epsilon)= E  F (u,v)\\
\\
\left[ - \frac{\hbar^2}{2m} \frac{\partial^2}{\partial u^2} + V (u) \right] F (u,v) - \frac{\hbar^2}{2m} \delta' (u-a)  F (u +\epsilon,v+\epsilon)
+\frac{\hbar^2}{2m} \delta' (u-b) F (u -\epsilon,v-\epsilon)= E F (u,v),
\end{array} \right.
\end{equation}
from where it follows that $F(u,v)=\xi(u)\psi(v)$ and so $\tilde F(x,y)=\xi(x-y)\psi(x+y)$.
Furthermore, since $F^W(x,p)$ is real, eq.(42) implies that $\xi=\psi^*$ and so $\tilde F(x,y)=\psi^*(x-y)\psi(x+y)$, where the complex function $\psi$ obeys the equation:
\begin{equation}
-\frac{\hbar^2}{2m} \psi'' (x) + V(x) \psi (x) -\frac{\hbar^2}{2m} \delta' (x-a) \psi (x^+) +\frac{\hbar^2}{2m} \delta' (x-b) \psi (x^- ) = E \psi (x),
\end{equation}
where we performed the limit $\epsilon \to 0^+$.
Finally, if $F^W (x,p) $ satisfies integral Dirichlet conditions at $x=a^+$ and $x=b^-$ then the corresponding wave function satisfies $\psi(a^+)=\psi(b^-)=0$, eq.(19).

In summary, we have shown that if $F^W$ satisfies the conditions (i) to (iv) above then $F^W$
is the Wigner transform (2) of a complex function $\psi$ which 1) is a solution of the Schr\"odinger equation (46) and 2) obeys Dirichlet boundary condition at $x=a^+$ and $x=b^-$.
In section 6 we will show how to solve this equation and prove that its unique (up to a global phase factor) solution is the standard solution of the corresponding eigenvalue problem in operator quantum mechanics.

\section{Boundary dynamics}

Let us now consider the time evolution:
\begin{equation}
\frac{\partial F^W}{\partial t}= \Theta_1 \frac{\partial F^W_1}{\partial t}+ \Theta_2 \frac{\partial F^W_2}{\partial t},
\end{equation}
and focus on the time derivative $\partial F^W_1 / \partial t$:
\begin{equation}
\begin{array}{c}
\frac{\partial F^W_1}{\partial t} = \frac{1}{\pi \hbar} \int_{a-x}^{x-a} dy e^{- 2i p y/ \hbar} \left[ \frac{\partial \psi^*}{\partial t} (x-y) \psi (x+y) + \psi^* (x-y) \frac{\partial \psi}{\partial t} (x+y) \right] =\\
\\
= \frac{1}{i \pi \hbar^2} \int_{a-x}^{x-a} dy e^{- 2i p y/ \hbar} \left\{\frac{\hbar^2}{2m} \left[ \psi''^* (x-y) \psi (x+y) - \psi^* (x-y) \psi'' (x+y) \right] + \right.\\
\\
\left. +  \left[V (x+y) - V(x-y ) \right] \psi^* (x-y) \psi (x+y)  \right\},
\end{array}
\end{equation}
where we used the Schr\"odinger equation. Consider the first term. After a few integrations by parts and by keeping track of the boundary contributions and Dirichlet's condition, we obtain:
\begin{equation}
\begin{array}{c}
\frac{1}{2 i m \pi } \int_{a-x}^{x-a} dy e^{- 2i p y/ \hbar} \psi''^* (x-y) \psi (x+y) = \frac{i}{2  m \pi } \left\{e^{- \frac{2ip}{\hbar} (x-a)} \psi'^* (a^+) \psi (2x -a^-) - \right.\\
\\
\left. -e^{- \frac{2ip}{\hbar} (a-x)} \psi^* (2x -a^-) \psi' (a^+) \right\}
  + \frac{1}{2 i m  \pi } \int_{a-x}^{x-a} dy e^{- 2i p y/ \hbar} \psi^* (x-y) \psi'' (x+y) - \frac{p}{m}\frac{\partial F^W_1}{\partial x}.
\end{array}
\end{equation}
The penultimate term exactly cancels the second term on the right-hand side of eq.(48). The first two terms in eq.(49) can be written as:
\begin{equation}
\begin{array}{c}
-\frac{ \hbar}{4  m \pi } \int_{- \infty}^{+ \infty} dk \left[e^{ik (a-x)} + e^{ik (x-a)}\right] k F^W_1 \left( x^+, p - \frac{\hbar}{2}k \right) = \\
\\
= - \frac{i  \hbar}{2  m } \left[ F^W_1 (x^+,p) * \delta' (x-a) - \delta' (x-a)  * F^W_1 (x^+,p) \right] = - \frac{\hbar^2}{2m} \left[ \delta' (x-a) , F^W_1 (x^+,p) \right]_M.
\end{array}
\end{equation}
The terms involving the potential $V(x)$ are familiar in Wigner quantum mechanics. They yield the contribution $\left[ V(x), F^W_1 (x,p) \right]_M$. After assembling all the results, we get:
\begin{equation}
\frac{\partial F^W_1}{\partial t} (x,p;t)  = \left[\frac{p^2}{2m}+V(x), F^W_1 (x,p;t) \right]_M - \frac{\hbar^2}{2m} \left[\delta'(x-a), F^W_1 (x^+,p;t) \right]_M .
\end{equation}
A similar calculation leads to:
\begin{equation}
\frac{\partial F^W_2}{\partial t} (x,p;t)  = \left[\frac{p^2}{2m}+V(x), F^W_2 (x,p;t) \right]_M + \frac{\hbar^2}{2m} \left[\delta'(x-b), F^W_2 (x^-,p;t) \right]_M.
\end{equation}
From these results we get for $F^W(x,p;t) = \Theta_1(x) F^W_1 (x,p;t) + \Theta_2(x) F^W_2 (x,p;t)$ :
\begin{eqnarray}
\frac{\partial F^W}{\partial t} (x,p;t)  = \left[\frac{p^2}{2m}+V(x), F^W (x,p;t) \right]_M
&-& \frac{\hbar^2}{2m} \left[\delta'(x-a), F^W (x^+,p;t) \right]_M \nonumber \\
&+& \frac{\hbar^2}{2m} \left[\delta'(x-b), F^W (x^-,p;t) \right]_M,
\end{eqnarray}
where we used the set of relations given by eqs.(33-36).

A straightforward consequence is that probability is conserved (as expected):
\begin{equation}
\frac{\partial}{\partial t} \int_a^b dx \int_{- \infty}^{+ \infty} dp F^W (x,p;t) = \frac{\partial}{\partial t} \int_{- \infty}^{+ \infty} dx \int_{- \infty}^{+ \infty} dp \left[ \Theta_1 (x) F^W_1 (x,p;t) + \Theta_2 (x) F^W_2 (x,p;t) \right] =0.
\end{equation}

\section{Weyl transform of the standard quantum description}

The standard approach to derive the Wigner formulation of the eigenvalue problem for confined systems would amount to applying the Weyl transform, $W$ eq.(1), to both the eigenvalue equation and to the boundary conditions. However, we saw in section 2, that this procedure does not yield the correct results. As we shall see, this is linked to the way operator quantum mechanics handles confined dynamical systems. The standard procedure is to 1) solve the eigenvalue equation for the unconfined wave function $\phi(x)$ and 2) impose the wave function to be zero outside the bulk and produce the confined state $\psi(x)=\phi(x)\theta(x-a)\theta(b-x)$. The problem in the Wigner formulation is that $W(|\psi><\psi|)\not= W(|\phi><\phi|)\theta(x-a) \theta(b-x)$ and thus we cannot solve the unconfined stargenvalue equation and manipulate its solution to obtain the confined Wigner function.

The reason for this apparent contradiction between the standard operator and Wigner formulations of quantum mechanics resides in the fact that $\psi$ is not a globally valid solution of the operator eigenvalue equation. Globally, $\psi$ satisfies a different eigenvalue equation which includes boundary potentials. Given the intrinsically non-local character of the Wigner function this fact is crucial to derive the correct Wigner description of the bounded system.

In this section we will a) derive a globally valid eigenvalue equation for confined systems satisfying Dirichlet boundary conditions, b) show that the only solution of this equation is the confined wave function and c) shortly discuss the numerical implementation of the new eigenvalue equation. It will then be trivial to realize that the Weyl transform of the new eigenvalue equation is the bounded stargenvalue equation (37,38), thus recovering the usual correspondence between standard operator and Wigner quantum mechanics.

\subsection{Bounded eigenvalue equation}

Let then $\phi$ be the unconfined solution of the eigenvalue equation satisfying Dirichlet boundary conditions at $x=a$ and $x=b$:
\begin{equation}
\hat H \phi = E \phi \quad , \quad \hat H= \frac{\hat p^2}{2m} + V(\hat x) \quad \mbox{and} \quad \phi(a)=\phi(b)=0 .
\end{equation}
Notice that $\phi (x)$ is the solution of eq.(55) everywhere, i.e. for all $x\in {\cal R}$. To obtain the confined wave function (for instance for a particle in a box) the standard procedure is to cut off $\phi$ outside the box and produce the new wave function: $\psi(x)=\phi(x)\theta(x-a)\theta(b-x)$. Notice that $\psi$ satisfies eq.(55) inside the box, it also satisfies it outside the box, but it does not satisfy it at the boundaries. Hence, $\psi$ is not a global solution of (55). Due to non-local effects the Weyl transform of $\psi$ does not satisfy the stargenvalue equation (6), even in the bulk.

It seems quite natural to expect that if we are able to provide the global eigenvalue equation for $\psi$ then the correct stargenvalue equation would just be its Weyl transform. The aim of this section is to derive this global equation. The first step is to introduce a twice differentiable version of the Heaviside step function and of the confined wave function $\psi$:
\begin{equation}
\theta_{\epsilon}(x):
\left\{ \begin{array}{l}
\in C^2({\cal R}) \quad \mbox{and is non decreasing} \\
\\
=0, \quad x< -\epsilon   \hspace{5cm} , \qquad \psi _{\epsilon}(x) =  \phi(x) \theta _{\epsilon}(x-a) \theta _{\epsilon}(b-x). \\
\\
=1, \quad x > \epsilon \quad ,\quad 0<\epsilon <<1
\end{array} \right. \\
\end{equation}
The definition of $\theta_{\epsilon}(x)$ is fully compatible with the distribution $\theta(x)$ we used in the previous sections: $\lim_{\epsilon \to 0^+} \theta_{\epsilon}(x)=0$ if $x<0$,  $\lim_{\epsilon \to 0^+} \theta_{\epsilon}(x)=1$ if $x>0$ and $\lim_{\epsilon \to 0^+} \theta_{\epsilon}(0)$ remains unspecified (it might be any number between $0$ and $1$ or it may not be defined at all). The results of this section are not dependent of the particular function $\theta_{\epsilon}(x)$ but only of the general properties given in (56).
From (56) the smooth versions of $\delta(x)$ and $\delta'(x)$ follow immediately:
$\delta_{\epsilon}(x)=\theta'_{\epsilon}(x)$ and $\delta_{\epsilon}'(x)=\theta''_{\epsilon}(x)$. It is also clear that $\lim_{\epsilon \to 0^+} (\delta_{\epsilon}(x),\delta'_{\epsilon}(x),\psi_{\epsilon} (x)) =
(\delta(x),\delta'(x),\psi (x))$.
Let us then apply $\hat H$ to $\psi_{\epsilon} (x)$:
\begin{eqnarray}
\hat H \psi_{\epsilon} (x) & = & \left\{ -\frac{\hbar^2}{2m} \frac{\partial ^2}{\partial x^2} + V(x) \right\} \psi_{\epsilon}(x) \nonumber \\
& =& \left\{ -\frac{\hbar^2}{2m} \phi''(x) + V(x)\phi(x) \right\}
\theta_{\epsilon}(x-a) \theta_{\epsilon}(b-x) \nonumber \\
& - & \frac{\hbar^2}{2m} \left\{ 2 \phi'(x) \delta_{\epsilon} (x-a) -2 \phi'(x) \delta_{\epsilon} (b-x) + \phi(x) \delta'_{\epsilon} (x-a) +\phi(x) \delta_{\epsilon}' (b-x) \right\} \nonumber \\
&=& E \psi_{\epsilon} (x)+ \frac{\hbar^2}{2m} \left\{\phi(x) \delta_{\epsilon}' (x-a) - \phi(x) \delta_{\epsilon}' (x-b) \right\},
\end{eqnarray}
where in the last step we use the fact that $\phi(x)$ satisfies the eigenvalue equation (55) and that the Dirichlet boundary conditions on $\phi(x)$ imply for sufficiently small $\epsilon$: $\phi'(x) \delta_{\epsilon} (x-a) =-\phi(x) \delta_{\epsilon}' (x-a)$ and likewise $\phi'(x) \delta_{\epsilon} (b-x) =-\phi(x) \delta_{\epsilon}' (x-b)$. The former equation is not yet a closed eigenvalue equation for $\psi_{\epsilon} (x)$ since it also evolves the wave function $\phi(x)$.

In what follows and to simplify the discussion we will consider the case of just one boundary placed at $a=0$. The derivation of the two-boundary eigenvalue equation follows exactly the same steps and will be written explicitly at the end.
In the one boundary case eq.(57) reduces to:
\begin{equation}
\hat H \psi_{\epsilon} (x)= E \psi_{\epsilon} (x) + \frac{\hbar^2}{2m} \delta_{\epsilon}' (x) \phi(x),
\end{equation}
where $\phi(x)$ satisfies eq.(55) with Dirichlet boundary conditions at $x=0$. The former equation follows from applying $\hat H$ to the state $\psi_{\epsilon} (x)=\phi(x) \theta_{\epsilon} (x)$.

To proceed we notice that for $x>\epsilon$ the two equations (55,58) - for $\phi(x)$ and for $\psi_{\epsilon}(x)$, respectively - are identical. If we supply identical boundary conditions, for instance, at $x=2\epsilon$ we will get identical solutions for $x>\epsilon$, i.e. if $\phi(2\epsilon)=\psi_{\epsilon}(2\epsilon)$ and $\phi'(2\epsilon)=\psi'_{\epsilon}(2\epsilon)$ then $\phi(x)=\psi_{\epsilon}(x)$ for all $x>\epsilon$. Furthermore, and since $\phi(x)$ is analytical, we also have:
\begin{equation}
\phi(x)=\phi(x+2\epsilon) + \sum_{n=1}^{\infty} \frac{(-2\epsilon)^n}{n!} \frac{\partial^n \phi}{\partial x^n} (x+2\epsilon),
\end{equation}
and for $x>-\epsilon$ we have $\phi(x+2\epsilon)=\psi_{\epsilon}(x+2\epsilon)$ and equally $
\frac{\partial^n \phi}{\partial x^n} (x+2\epsilon)=\frac{\partial^n \psi_{\epsilon}}{\partial x^n} (x+2\epsilon)$, $\forall n \in {\cal N}$. Hence:
\begin{equation}
\phi(x)=\psi_{\epsilon}(x+2\epsilon) + \sum_{n=1}^{\infty} \frac{(-2\epsilon)^n}{n!} \frac{\partial^n \psi_{\epsilon}}{\partial x^n} (x+2\epsilon) \quad , \quad \forall x>-\epsilon .
\end{equation}
The trick here is that while $\phi(x)$ is analytical $\psi_{\epsilon}(x)$ is not (cf.(56)) and thus the Taylor expansion (60) yields $\phi(x)$ instead of $\psi_{\epsilon}(x)$. Substituting the former expansion in eq.(58) we get:
\begin{equation}
\hat H \psi_{\epsilon} (x)= E \psi_{\epsilon} (x) + \frac{\hbar^2}{2m} \delta'_{\epsilon} (x) \left\{ \psi_{\epsilon}(x+2\epsilon) + \sum_{n=1}^{\infty} \frac{(-2\epsilon)^n}{n!} \frac{\partial^n \psi_{\epsilon}}{\partial x^n} (x+2\epsilon) \right\}.
\end{equation}
Notice that the previous equation is valid for all $x\in {\cal R}$ because $\delta'_{\epsilon} (x)=0$ if $x < -\epsilon $. Finally by taking the limit $\epsilon \to 0^+$ we obtain:
\begin{equation}
\hat H \psi (x)= E \psi (x) + \frac{\hbar^2}{2m} \delta' (x) \psi(x^+),
\end{equation}
where we used the notation $x^+$ to make it explicit that in the product of $\delta'$ by $\psi$ the two factors are evaluated at different points. This is crucial because equation (61) is valid for all $x\in{\cal R} $ only in this case, i.e. in the limit $\epsilon  \to 0^+$ but not for $\epsilon=0$. Equation (62) is the new eigenvalue equation for $\psi(x)$. Following exactly the same procedure we can easily generalize it to the case of a double boundary placed at $x=a$ and $x=b$, $(a<b)$. We get:
\begin{equation}
\hat H \psi (x)= E \psi (x) + \frac{\hbar^2}{2m} \left\{\delta' (x-a) \psi(x^+)-
\delta' (x-b) \psi(x^-) \right\},
\end{equation}
which together with the boundary conditions $\psi(a^+)=\psi(b^-)=0$ provide the operator formulation of the confined eigenvalue problem for Dirichlet boundary conditions.
To finish let us point out that: 1) eq.(63) was obtained in section 4 from the stargenvalue equations (37,38) and 2) conversely, the Weyl transform of eq.(63) immediately yields the stargenvalue equations (37,38).

\subsection{Solving the new eigenvalue equation}

Our next step is to solve the modified eigenvalue equation (63) and show that the confined wave function $\psi(x)$ is its only solution. To make it simple let us consider again the one boundary example eq.(62).

For $x>0$, $\delta(x)=0$ and eq.(62) reduces to the unconfined equation (55). The boundary conditions $\psi(0^+)=0$ and $\psi'(0^+)=\phi'(0^+)$ impose $\psi(x)=\phi(x)$ for all $x>0$.

For $-\sigma \le x \le \sigma$ and $\sigma <<1$ we have:
\begin{eqnarray}
& & \int_{-\sigma}^{x} \hat H \psi (x') dx' = \int_{-\sigma}^{x} E \psi(x') dx' +
\frac{\hbar^2}{2m} \int_{-\sigma}^{x} \delta' (x') \psi(x'+0^+) dx' \nonumber \\
& \Longrightarrow & -\frac{\hbar^2}{2m} \left\{ \psi'(x)-\psi'(-\sigma )\right\}
+ {\cal O} (x+\sigma) =\nonumber \\
& =& E {\cal O} (x+\sigma) +
\frac{\hbar^2}{2m} \left\{ \left[ \delta(x')\psi(x'+0^+) \right]_{-\sigma}^{x} - \int_{-\sigma}^{x} \delta (x') \psi'(x'+0^+) dx' \right\} \nonumber \\
& \Longrightarrow & \psi'(x)-\psi'(-\sigma) = \psi'(0^+) \theta(x) + {\cal O} (x+\sigma),
\end{eqnarray}
where in the first step we introduced the notation ${\cal O}(y)$ to designate an arbitrary continuous function such that $\lim_{y \to 0} {\cal O}(y)=0$, and in the third step we used the fact that $\psi(0^+)=0$.
Taking the limits $\sigma \to 0^+$ and $x \to 0^+$ we get:
\begin{equation}
\psi'(0^+)-\psi'(0^-) = \psi'(0^+) \Longrightarrow \psi'(0^-)=0.
\end{equation}
The second integration of the eigenvalue equation yields:
\begin{eqnarray}
\int_{-\sigma}^{\sigma} \psi'(x) dx - \int_{-\sigma}^{\sigma} \psi'(-\sigma) dx &=& \int_{-\sigma}^{\sigma} \psi'(0^+) \theta (x) dx + \int_{-\sigma}^{\sigma} {\cal O} (x+\sigma) dx \nonumber \\
\Longrightarrow \psi (\sigma)- \psi(-\sigma) & = & \sigma \left\{2 \psi'(-\sigma) + \psi'(0^+) \right\} + {\cal O} (\sigma^2),
\end{eqnarray}
and in the limit $\sigma \to 0^+$ we have $\psi(0^+)-\psi(0^-)=0$, which together with the original Dirichlet boundary conditions at $x=0^+$ imply $\psi(0^-)=0$.

We conclude that for $x<0$, $\psi(x)$ satisfies eq.(55) with boundary conditions $\psi(0^-)=\psi'(0^-)=0$ and thus $\psi(x)=0$, $\forall x<0$. Hence, we are left with the original bounded wave function: $\psi(x) = \phi (x) \theta (x)$. The generalization for a system with two boundaries is straightforward. Following exactly the same procedure we solve eq.(63)
and get $\psi(x)=\phi(x) \theta(x-a) \theta(b-x)$, as we should. This result concludes the proof of Backer's converse construction initiated in section 4.

\subsection{The free particle bounded eigenvalue equation: numerical solution of a simple example}

As a simple example let us consider the case of a free particle confined to the interval $]-1,1[$ and subject to Dirichlet boundary conditions.

The unconfined Hamiltonian in the position representation is given by $\hat H = -\frac{\hbar^2}{2m} \frac{\partial^2}{\partial x^2}$ and the unconfined fundamental state (solution of eq.(55)) is: $\phi_1(x)=\cos (\frac{\pi}{2}x)$ with $E=E_1=\hbar^2\pi^2/8m$. Fig. 1 displays the unconfined wave function for $\hbar=m=1$.

We now consider a first possible approximation to the confined
eigenvalue equation (63). We make $\epsilon=0.25$ and use a Gaussian
approximation to the Dirac delta function: $\tilde
\delta_{\epsilon}(x) = \frac{1}{\pi^{1/2} \epsilon'} \exp
\left\{-(\frac{x}{\epsilon'})^2\right\}$ where $\epsilon'$
substitutes $ \epsilon$ for the following reason: the function
$\tilde \delta_{\epsilon}(x)$ is not identically zero for
$x>\epsilon$ nor for $x< -\epsilon$ and thus there is no function
$\theta_{\epsilon}(x)$ satisfying the conditions (56) and such
that $\tilde \delta_{\epsilon}(x) =\theta'_{\epsilon}(x)$.
However, if we define $\tilde \delta_{\epsilon}(x)$ using a
sufficiently small spread $\epsilon'$ (when compared with
$\epsilon$) then, for our numerical purposes $\tilde \delta_{\epsilon}(x)$
will provide a good
enough approximation to $\theta'_{\epsilon}(x)$.  In all future
applications we will make: $\epsilon'=0.8\epsilon$. Furthermore,
and to simplify the notation we will refer to $\tilde
\delta_{\epsilon}(x) $ as just $\delta_{\epsilon}(x) $. Finally,
taking into account the considerable large value of $\epsilon $ we
have to consider the contribution of higher order terms in the
expansion (60), and thus we will use the two boundaries
generalization of eq.(61) as our starting point and truncate the
series at the third term. Hence, the equation we want to solve is:
\begin{equation}
\begin{array}{c}
-\frac{\hbar^2}{2m} \frac{\partial^2}{\partial x^2} \psi_{\epsilon} (x)= E \psi_{\epsilon} (x) + \frac{\hbar^2}{2m} \delta_{\epsilon}' (x+1) \left\{
\psi_{\epsilon}(x+2 \epsilon)-2\epsilon \psi_{\epsilon}'(x+2 \epsilon)
+2\epsilon^2 \psi_{\epsilon}''(x+2 \epsilon)\right\}\\
\\
-\frac{\hbar^2}{2m} \delta_{\epsilon}' (x-1) \left\{ \psi_{\epsilon}(x-2 \epsilon)
+2\epsilon \psi_{\epsilon}'(x-2 \epsilon)
+2\epsilon^2 \psi_{\epsilon}''(x-2 \epsilon) \right\}.
\end{array}
\end{equation}
Fig. 2 displays the numerical solution of the former equation for the boundary conditions $\psi_{\epsilon}(a+2\epsilon=-0.5)=\psi_{\epsilon}(b-2\epsilon=0.5)=\phi(0.5)=0.707$ and for $E=E_1=\pi^2/8$, $\hbar=m=1$. Notice that the wave function is not completely confined which follows from the fact that we are not using the true Dirac delta but a (poor) Gaussian approximation of it. If we decrease the value of $\epsilon$ the confinement will increase greatly.

Let us then consider a second approximation to the confined eigenvalue equation by making $\epsilon=0.0025$. Given the considerable small value of $\epsilon$ we can now truncate the series in the two boundaries generalization of eq.(61) to zero order and numerically solve the finite version of eq.(63):
\begin{equation}
-\frac{\hbar^2}{2m} \frac{\partial^2}{\partial x^2} \psi_{\epsilon} (x)= E \psi_{\epsilon} (x) + \frac{\hbar^2}{2m} \left\{ \delta_{\epsilon}' (x+1)
\psi_{\epsilon}(x+2 \epsilon) -\delta_{\epsilon}' (x-1) \psi_{\epsilon}(x-2 \epsilon)\right\}.
\end{equation}
Fig. 3 displays the numerical solution of the former equation for the boundary conditions $\psi_{\epsilon}(a+2\epsilon=-0.995)=\psi_{\epsilon}(b-2\epsilon=0.995)=\phi(0.995)=0.008$ and $\hbar=m=1$, $E=E_1=\pi^2/8$.

We will see in the next section that the part of the wave function that escapes from the box corresponds to quantum trajectories with energies much higher than the fundamental state. By making $\epsilon \to 0^+$ we completely confine the wave function.

\section{Wigner trajectories in confined systems}

In this section, we briefly review the topic of Wigner trajectories \cite{Lee1,Lee2,Lee3,Lee4}, where the previous work may find some interesting applications. There are two well-known instances, where this concept plays an important role: collision processes \cite{Lee1,Carruthers} and the study of quantum systems that exhibit chaotic behavior in the classical limit \cite{Lee1,Takahashi,Latka,Shin,Lin}.

The Schr\"odinger and Wigner formulations of a collision process are obviously equivalent. This notwithstanding, there are important practical differences between the two approaches. In most cases, where an exact solution to the problem is unknown, one has to devise some approximative scheme. If certain conditions hold (concerning the interaction potential), the Moyal equation (5) can be truncated to a good degree of accuracy. Since it does not contain any operator, it is often easier do develop approximations on it than on the Schr\"odinger equation. Having computed the Wigner function that solves this truncated version of the Moyal equation, one can subsequently evaluate quantum corrections to the classical phase space trajectories. In addition to the benefit of this pictorial and intuitive description in terms of trajectories there is another advantage in this approach: collision cross sections can be computed by applying the same numerical techniques used in classical collision processes.

Another situation where the Wigner trajectories play an important role is that of nonlinear systems exhibiting classical chaotic behavior.
In these systems the classical phase space trajectories are of random nature while the Wigner trajectories are smooth paths thus showing that the chaotic behavior disappears upon quantization. A topic of current interest is that of understanding the transition from classical to quantum dynamics. Since the Wigner trajectories might be computed as an $\hbar$ order by order correction to the classical trajectories they provide a powerful, pictorial tool to study this transition.

In this section we will a) shortly review the subject of Wigner trajectories for unconfined systems, b) discuss the main difficulties in extending the formalism to bounded systems and show that our previous results provide a new approach in this context and c) illustrate these results using the simple example of a free particle confined to an infinite square well. The reader should be advised that our aim is not to provide a fully developed and precise formulation of these topics but only to outline (and hopefully motivate) some possible lines of research.

\subsection{Wigner trajectories}

Let us then consider the Moyal equation of motion (5) associated to $H=\frac{p^2}{2m}+V(x)$:
\begin{equation}
\frac{\partial F^W}{\partial t} (x,p,t) = - \frac{p}{m} \frac{\partial F^W}{\partial x} + \frac{\partial V}{\partial x} \frac{\partial F^W}{\partial p} + \frac{\left( \hbar / 2i
\right)^2}{3!} \frac{\partial^3 V}{\partial x^3} \frac{\partial^3 F^W}{\partial p^3} + \cdots,
\end{equation}
Alternatively, we can write this equation as:
\begin{equation}
\frac{\partial F^W}{\partial t} (x,p,t) = - \frac{p}{m} \frac{\partial F^W}{\partial x} +
\int_{- \infty}^{+ \infty} d p' \hspace{0.3 cm} J (x, p') F^W (x, p + p' ,t),
\end{equation}
where
\begin{equation}
J(x,p) = \frac{i}{\pi \hbar^2} \int_{- \infty}^{+ \infty} d y \hspace{0.3 cm} \left[ V (x+y) - V (x-y) \right] e^{-2i p y / \hbar}.
\end{equation}
In the absence of a boundary and if the potential $V$ is at most quadratic in $x$, then the Moyal equation (69) reduces to the Liouville equation:
\begin{equation}
\frac{\partial F^W}{\partial t} (x,p,t) = - \frac{p}{m} \frac{\partial F^W}{\partial x} + \frac{\partial V}{\partial x} \frac{\partial F^W}{\partial p} .
\end{equation}
The previous equation means that the solution $F^W (x,p,t)$ evolves along the classical trajectories, i.e.:
\begin{equation}
\left\{
\begin{array}{l}
F^W (x,p,t) = F^W \left( x(-t), p(-t), 0 \right),\\
\\
\dot x = \frac{p}{m}, \qquad \dot p = - \frac{\partial V}{\partial x}.
\end{array}
\right.
\end{equation}
This motivates the concept of Wigner trajectories. If we define an effective quantum potential $V_{eff} (x,p,t)$ according to:
\begin{equation}
\frac{\partial V_{eff}}{\partial x} \frac{\partial F^W}{\partial p} = \int_{- \infty}^{+ \infty} d p' \hspace{0.3 cm} F^W (x,p + p' ,t) J(x,p'),
\end{equation}
then we can express the Moyal equation in the form:
\begin{equation}
\frac{\partial F^W}{\partial t}  = - \frac{p}{m} \frac{\partial F^W}{\partial x} + \frac{\partial V_{eff}}{\partial x} \frac{\partial F^W}{\partial p} ,
\end{equation}
which can be interpreted in analogy with (72), by stating that the Wigner function evolves along the Wigner trajectories given by:
\begin{equation}
\dot x = \frac{p}{m}, \qquad \dot p = - \frac{\partial V_{eff}}{\partial x}.
\end{equation}
If an exact solution for $F^W$ is known, then we can, in principle, calculate the effective potential from (74) and subsequently solve (76) to obtain the Wigner trajectories. However, in general, such is not the case. If the potential $V(x)$ does not deviate appreciably from a quadratic potential then we can address the problem by iteratively evaluating order by order the quantum corrections to the classical trajectories \cite{Lee1,Lee3,Lee4}. The zero order approximation (i.e. the classical solution), satisfies the Liouville equation:
\begin{equation}
\frac{\partial F^W_0}{\partial t}  = - \frac{p}{m} \frac{\partial F^W_0}{\partial x} + \frac{\partial V}{\partial x} \frac{\partial F^W_0}{\partial p} .
\end{equation}
The order zero trajectories correspond to the classical Hamilton equations. The order one equation is given by:
\begin{equation}
\frac{\partial F^W_1}{\partial t}  = - \frac{p}{m} \frac{\partial F^W_1}{\partial x} + \frac{\partial V}{\partial x} \frac{\partial F^W_1}{\partial p} + \frac{ \left( \hbar / 2i \right)^2}{3!} \frac{\partial^3 V}{\partial x^3} \frac{\partial^3 F^W_0}{\partial p^3} .
\end{equation}
If we solve this equation for $F_1^W$ and substitute in (75,76), we obtain the first order trajectories:
\begin{equation}
\left\{
\begin{array}{l}
\dot x = \frac{p}{m},\\
\\
\dot p = - \frac{\partial V}{\partial x} + \frac{1}{3!} \left( \frac{\hbar}{2i} \right)^2 \frac{\partial^3 V}{\partial x^3} \frac{\partial^3 F^W_0}{\partial p^3} / \frac{\partial F^W_1}{\partial p}.
\end{array}
\right.
\end{equation}
The second order approximation is:
\begin{equation}
\frac{\partial F^W_2}{\partial t}  = - \frac{p}{m} \frac{\partial F^W_2}{\partial x} + \frac{\partial V}{\partial x} \frac{\partial F^W_2}{\partial p} + \frac{ \left( \hbar / 2i \right)^2}{3!} \frac{\partial^3 V}{\partial x^3} \frac{\partial^3 F^W_1}{\partial p^3} +
\frac{ \left( \hbar / 2i \right)^4}{5!} \frac{\partial^5 V}{\partial x^5} \frac{\partial^5 F^W_0}{\partial p^5},
\end{equation}
and yields the second order correction to the classical trajectories:
\begin{equation}
\left\{
\begin{array}{l}
\dot x = \frac{p}{m},\\
\\
\dot p = - \frac{\partial V}{\partial x} + \frac{1}{3!} \left( \frac{\hbar}{2i} \right)^2 \frac{\partial^3 V}{\partial x^3} \frac{\partial^3 F^W_1}{\partial p^3} / \frac{\partial F^W_2}{\partial p} + \frac{1}{5!} \left( \frac{\hbar}{2i} \right)^4 \frac{\partial^5 V}{\partial x^5} \frac{\partial^5 F^W_0}{\partial p^5} / \frac{\partial F^W_2}{\partial p}.
\end{array}
\right.
\end{equation}
This procedure can be carried on to any order in $\hbar$. Its main advantage is that it transforms the problem of solving an infinite order partial differential equation (the Moyal equation) into the problem of solving a sequence of first order partial differential equations. It also displays the advantage of casting the transition from classical to quantum dynamics as an order-by-order correction (in $\hbar$) to classical mechanics. There are, however, some weak points in the procedure, the most significant being perhaps that it is not always clear what is the degree of precision of an approximation to a given order nor if the entire scheme will converge.

\subsection{Wigner trajectories in bounded systems}

If a system has a boundary, then the former approximative scheme will obviously break down, because the boundary contribution was thus far unknown. As we have seen in the previous sections the boundary interaction has a non-local nature and will thus affect the Wigner trajectories well inside the bulk part of the system. It therefore must equally contribute to the quantum corrections of the trajectories.

In this case one was left with a single possible approach: to solve the Schr\"odinger equation for the confined wave function and compute the Wigner function from it. In some cases where an exact solution of the Schr\"odinger equation is known this method can be taken to completion. Take for instance the simple example of a free particle confined to the interval $]-1,1[$ and subject to Dirichlet boundary conditions \cite{Lee2}. The fundamental state is $\phi_1(x)=\cos(\frac{\pi}{2}x)$ from which we can compute the Wigner function eqs.(2,39,40). Since the particle is in an energy eigenstate we have:
\begin{equation}
\frac{\partial F^W}{\partial t} = [H,F^W]_M=0 \Longrightarrow F^W(q(-t),p(-t),0)=F^W(q,p,t)=F^W(q,p,0),
\end{equation}
and thus the Wigner function evolves along the paths where it displays a constant value, i.e. the Wigner trajectories $(q(-t),p(-t))$ are the equi-Wigner curves. Fig. 4 displays the Wigner trajectories for this simple system.

For more elaborate examples we are forced to use eq.(74) or eq.(75) to calculate the effective potential from the Wigner function and then eq.(76) to calculate the Wigner trajectories. For instance in \cite{Lee1,Lin} a system
in a symmetric box with Dirichlet boundary conditions and an external driving force was considered. The force, which is of the form $F(t) = F_0 \cos (\omega t)$, induces classically the appearance of resonance. Since the system has a boundary, quantum mechanically, the authors were compelled to work in the context of the Schr\"odinger equation but since an exact, analytic solution is not known the Wigner function and subsequently the Wigner trajectories could not be obtained exactly.

In general, the wave function based approach suffers from two major drawbacks: firstly, in some situations (as in the previous example) it is impossible to obtain an exact solution of the Schr\"{o}dinger equation which makes it difficult to infer what is the degree of precision of the resulting Wigner function. Secondly, even if an exact solution of the Schr\"{o}dinger equation is known, this method only provides the exact Wigner trajectories and in some situations it is desirable that the quantum corrections to the classical trajectories might be computed order by order in $\hbar$.

The results of this paper provide a new approach to determine the Wigner trajectories for bounded systems (satisfying Dirichlet boundary conditions) from an exclusive phase space (or c-number) point of view. In fact most of the formalism for the unconfined case can now be applied to the bounded case as well. This is certainly true for eq.(69-71,74-76) where one just has to take into account that: 1) $V(x)$ is no longer the bulk potential but it also encompasses a boundary contribution $V_D (x)$ which is associated with the Dirichlet boundary conditions and was given in eq.(53): $V_D (x)= \frac{\hbar^2}{2m} \left\{\delta'(x-b)-\delta'(x-a) \right\}$, 2) in the boundary corrections to eqs.(69,70,74) the Wigner function should be evaluated at $x^+$ and $x^-$ (in the left and right boundary correction, respectively).

It is not so clear if the approximative scheme (eq.(77) to eq.(81)) can be also extended to the bounded case. This is because the boundary potential deviates strongly from the harmonic oscillator potential and yields contributions to the Moyal equation to all orders in $\hbar$. We thus have to worry about the magnitude of these contributions and find a suitable criterion to determine the validity of the approximation. These issues are also a problem in the general unconfined case and are certainly not the subject of this paper. Our point here is just that the boundary potential approach is, thus far, the unique exact formulation of confined systems that only appeals to phase space objects and it thus seems to be the proper framework from which the former or other, more suitable, approximative scheme may be devised.

\subsection{Wigner trajectories of a particle in an infinite potential well}

In this section we ignore the issues of validity and convergence of the approximative scheme given by eqs.(77-81) and use it to compute the order by order quantum corrections to the classical trajectories of a free particle confined to the interval $]-1,1[$. Quite remarkably, we will see that this (in this context, very poor) approximation procedure is already able to reproduce most of the qualitative behavior of the exact quantum trajectories.

The exact Wigner trajectories for a particle in the fundamental state are solutions of eq.(76) with the effective potential satisfying (74,75) and were displayed in Fig. 4.
The zero order approximation to the Moyal equation (53) is given by:
\begin{equation}
\frac{\partial F_0^W}{\partial t} (x,p,t) = - \frac{p}{m} \frac{\partial F_0^W}{\partial x} + \frac{\hbar^2}{2m} \left\{ \frac{\partial V_R}{\partial x} \frac{\partial F_0^W}{\partial p} (x-\epsilon,p,t) -
\frac{\partial V_L}{\partial x} \frac{\partial F_0^W}{\partial p} (x+\epsilon,p,t) \right\},
\end{equation}
where $V_R(x)=\delta_{\epsilon}'(x-1)$ and $V_L(x)=\delta'_{\epsilon} (x+1)$. To simplify the discussion we assume that for sufficient small $\epsilon$ we have:
$$
\frac{\partial F_0^W /\partial p (x-\epsilon,p,t)}{\partial F_0^W /\partial p (x,p,t)}=
\frac{\partial F_0^W /\partial p (x+\epsilon,p,t)}{\partial F_0^W /\partial p (x,p,t)}= 1,
$$
and use these identities to transform eq.(83) into the Liouville equation (77) with $V=\frac{\hbar^2}{2m} (V_R-V_L)$.
The zero order approximation to the exact quantum trajectories are thus the solutions of eq.(73). They are displayed in Fig. 5.
where we used the Gaussian approximation to the Dirac delta function: $\delta_{\epsilon}(x)= e^{x^2/\epsilon'^2}/\epsilon' \sqrt{\pi}$, $\epsilon'=0.8 \epsilon$ and made $\epsilon=0.25$ (notice that this approximation was used in section 6 to compute the numerical solution of the bounded eigenvalue equation, fig. 2) and $m=\hbar=1$. For sufficient small $\epsilon$ these would be just the classical trajectories.

We now compute the first order corrections eq.(79). These are the solutions of:
\begin{equation}
\left\{
\begin{array}{l}
\dot x = \frac{p}{m},\\
\\
\dot p = \frac{\hbar^2}{2m} \left\{ \delta''_{\epsilon} (x+1) -\delta''_{\epsilon} (x-1) \right\}   + \frac{\hbar^2}{48 m} \left\{ \delta^{(4)}_{\epsilon} (x+1) - \delta^{(4)}_{\epsilon} (x-1)\right\},
\end{array}
\right.
\end{equation}
where we again ignored the contribution of the factors that involve the
(approximate) Wigner function. We get the trajectories displayed in fig. 6.

Finally, we compute the second order corrections (eq.(81)):
\begin{equation}
\left\{
\begin{array}{l}
\dot x = \frac{p}{m},\\
\\
\dot p = \frac{\hbar^2}{2m} \left\{ \delta''_{\epsilon} (x+1) -\delta''_{\epsilon} (x-1) \right\}   + \frac{\hbar^2}{48 m} \left\{ \delta^{(4)}_{\epsilon} (x+1) - \delta^{(4)}_{\epsilon} (x-1)\right\} -\\
\quad - \frac{\hbar^4}{3840 m} \left\{ \delta^{(6)}_{\epsilon} (x+1) - \delta^{(6)}_{\epsilon} (x-1)\right\},
\end{array}
\right.
\end{equation}
and get the trajectories displayed
in fig. 7.

Several remarks are now in order:\\
a) We see that the first order approximation already displays the deflection of the quantum trajectories towards regions of higher momentum as the particle approaches the boundaries.\\
b) In the second approximation the "islands" appear. They are not confined, as in the exact case, to regions of positive or negative momentum, but they are already confined to positive or negative positions. Furthermore, the deflection to regions of high momentum is now an exclusive feature (as in the exact case) of the higher momentum trajectories. The slower ones are essentially classical.\\
c) Not all Wigner trajectories will be confined. For sufficiently high momentum they will escape from the box. This is because we are not using the true Dirac delta but a Gaussian approximation to it. The free Wigner trajectories correspond to the part of the wave function that is outside from the box in fig. 2.\\
d) The former approximative scheme can be refined at least in three different directions: firstly by taking into account the contributions of the factors that are proportional to the Wigner function. Our, rather simple approximative scheme, is in fact, state independent and so could not possibly reproduce the exact Wigner trajectories. This step will be more demanding from the numerical point of view since one will be asked to solve the partial differential equation for the approximate Wigner function before computing each order correction to the classical trajectories. Secondly, by increasing the degree of approximation to the Dirac delta function, i.e. by decreasing the spread $\epsilon$. This is mandatory if one is interested in the behavior of the quantum trajectories of higher momentum. Lastly, and quite obviously by computing higher order corrections in $\hbar$.

\section{Conclusions}

In this paper we studied the Wigner-Weyl formulation of a particle
confined to a finite interval and subject to Dirichlet boundary
conditions. We found that: 1) the standard procedure of solving
the unconstrained system and then imposing the boundary conditions
does not work out in the Wigner-Weyl formulation; 2) the boundary
conditions and the confinement of the wave function determine a
boundary correction to the $*$-genvalue equation and to the Moyal
equation; 3) Dirichlet
boundary conditions on the confined wavefunction imply both
Dirichlet and Neumann boundary conditions on the corresponding
confined Wigner function.

Our main task was then the evaluation of the contributions of the
boundaries to the $*$-genvalue equation (6) and to the time
evolution equation (5) for a wave function satisfying Dirichlet
boundary conditions. These contributions were shown to have the
form of boundary potentials added to the Hamiltonian, and the same
potentials were obtained in the operator formulation of quantum
mechanics, when we derived the Schr\"odinger eigenvalue equation
for the confined wave function
$\psi(x)=\phi(x)\theta(x-a)\theta(b-x)$. Together with the
boundary conditions they are responsible, both in the Wigner and
in the Schr\"odinger formulations, for the confinement of the
system and the consequent discretization of the energy levels.
Finally, we extended Baker's converse construction to the bounded
case, proving the full equivalence between the
operator-Schr\"odinger and the Wigner formulation of the confined
eigenvalue problem.

Lastly, we used our previous results to approach the problem of evaluating
the Wigner trajectories of
bounded systems and ventured the possibility of applying these
methods to non-linear systems displaying classical chaotic
behavior.

The procedure described in this work can be generalized to
higher-dimensional multiparticle systems and also to any
$*$-genvalue equation of the form $ A^W (x,p) * F^W_a (x,p) = a
F^W_a (x,p)$, associated with the eigenvalue equation: $\hat A(
\hat x, \hat p ) \psi_a (x) =  a \psi_a (x)$. From a more general
point of view the results of this paper provide a first approach
to the problem of solving star-equations with boundaries. This is
a difficult but important problem as such equations are known to
play a key part in several fields of research ranging from
standard topics in quantum mechanics to some important
developments in M-theory \cite{pinzul}.

\vspace{1 cm}

\begin{center}

{\large{{\bf Acknowledgments}}}

\end{center}

\vspace{0.3 cm}
\noindent
We would like to thank Jo\~ao Marto, Jaime Prata and Aleksandar Mikovic for useful suggestions. This work was partially supported by the grants ESO/PRO/1258/98 and CERN/P/Fis/15190/1999.

\newpage

\section*{List of Figures}

\renewcommand{\baselinestretch}{1}
\small \normalsize

\renewcommand{\floatpagefraction}{0.7}
\renewcommand{\textfraction}{0}

\includegraphics[-3.5cm,16cm][1.5cm,19cm]{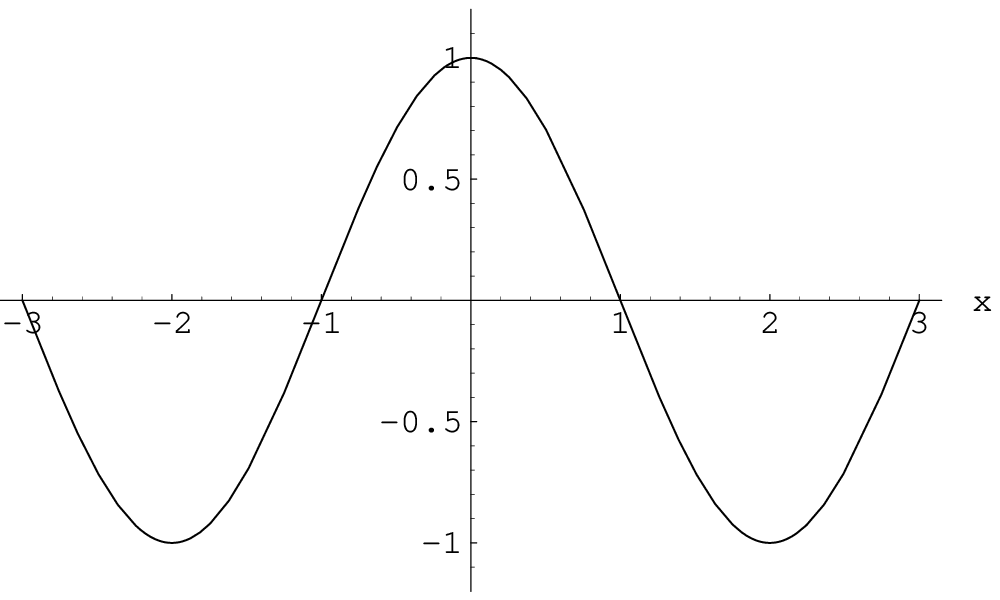}
\vspace{-3.3cm}
\begin{center}
$\phi_1(x)$
\end{center}
 \vspace{5.2cm}
\begin{center}
Fig.1: Unconfined fundamental solution of the free particle \\
eigenvalue equation satisfying Dirichlet boundary conditions \\
at $x=\pm 1$ and for $\hbar=m=1$
\end{center}
\vspace{-0.5cm}
\includegraphics[-3.5cm,21cm][1.5cm,24cm]{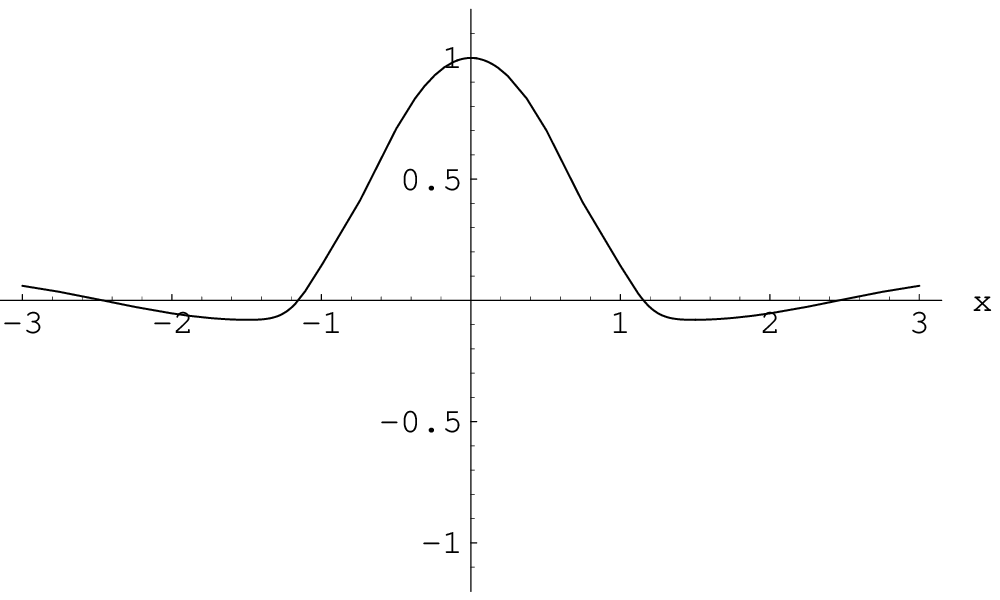}
\vspace{1.7cm}
\begin{center}
$\psi_{\epsilon}(x)$
\end{center}
 \vspace{5.2cm}
\begin{center}
Fig.2: Numerical solution of the first approximation (eq.67) to the \\
confined eigenvalue equation: $\epsilon=0.25$, $\hbar=m=1$ and $E=\pi^2/8$
\end{center}

\newpage
\renewcommand{\baselinestretch}{1}
\mbox{}
\includegraphics[-2.7cm,16cm][2.3cm,19cm]{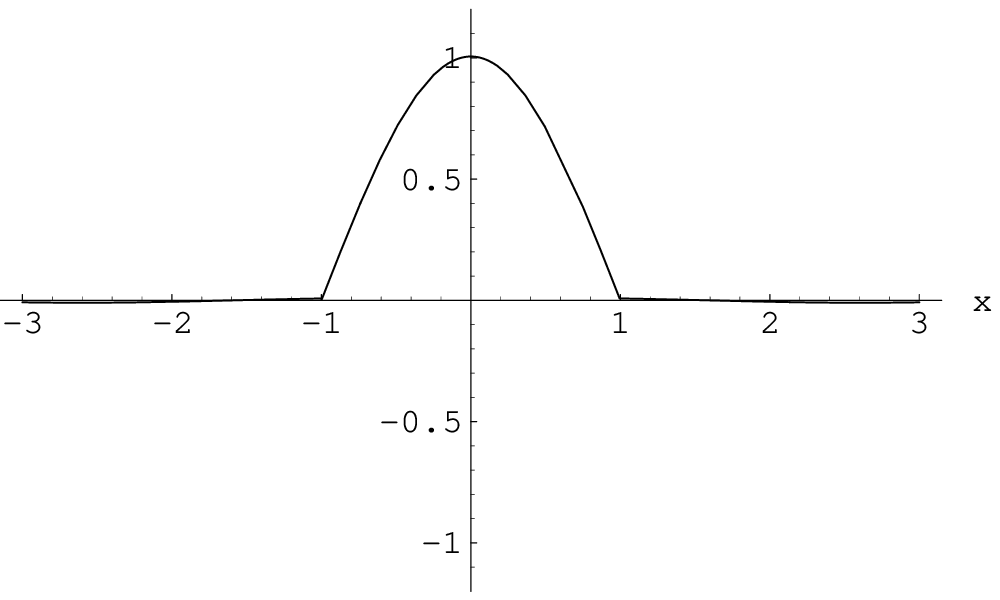}
\vspace{-3.3cm}
\begin{center}
$\psi_{\epsilon}(x)$
\end{center}
 \vspace{5.2cm}
\begin{center}
Fig.3: Numerical solution of the second approximation (eq.68) to the \\
confined eigenvalue equation: $\epsilon=0.0025$, $\hbar=m=1$ and
$E=\pi^2/8$
\end{center}
\vspace{2cm}
\includegraphics[-3.2cm,21cm][1.8cm,24cm]{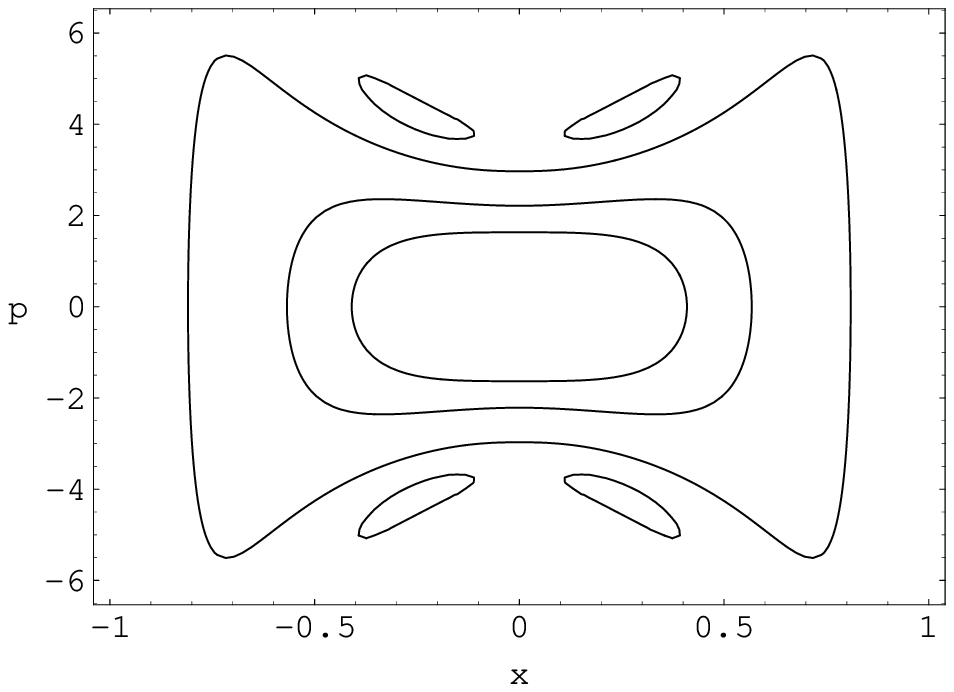}

\vspace{6.5cm}
\begin{center}
Fig.4: Exact Wigner trajectories for a particle in the \\
fundamental state and confined to the interval $]-1,1[$.
\end{center}

\newpage
\renewcommand{\baselinestretch}{1}
\mbox{}
\includegraphics[-2.8cm,19cm][2.2cm,22cm]{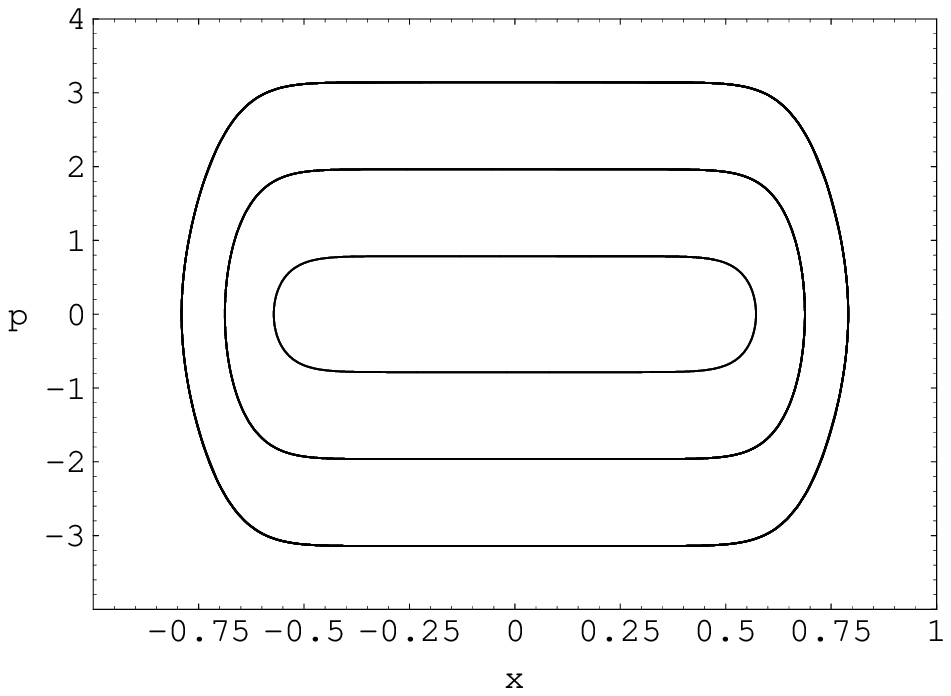}

 \vspace{4.3cm}
\begin{center}
Fig.5: Zero order or classical approximation to the \\
Wigner quantum trajectories: $\epsilon=0.25$ and $\hbar=m=1$.
\end{center}
\vspace{1.5cm}
%\end{minipage}
 %\vfill
%\begin{minipage}[t]{5.0cm}
\includegraphics[-3.5cm,23cm][1.5cm,24cm]{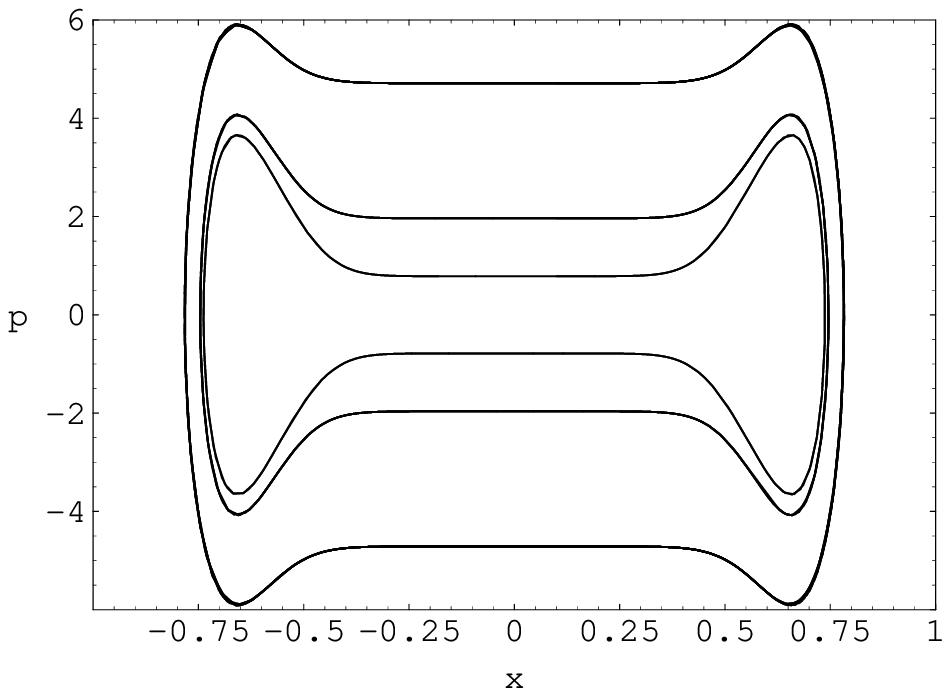}

 \vspace{8.2cm}
\begin{center}
Fig.6: First order approximation to the Wigner quantum \\
trajectories: numerical solution of eq.(84) with $\epsilon=0.25$
and $\hbar=m=1$.
\end{center}

\newpage
\renewcommand{\baselinestretch}{1}
\mbox{}
\includegraphics[-2.8cm,18.5cm][2.2cm,21.5cm]{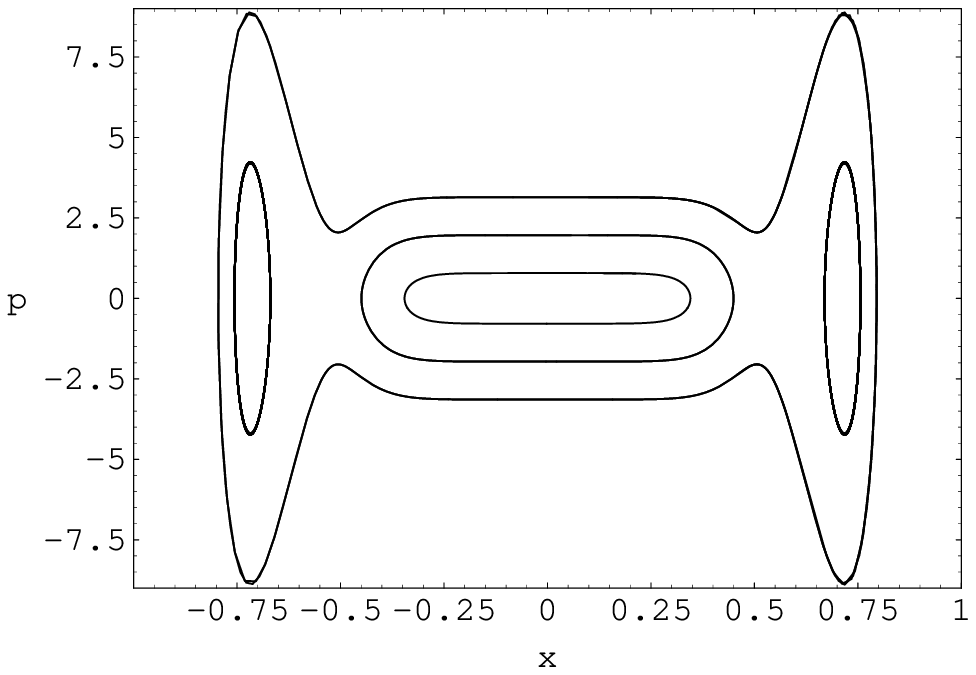}

 \vspace{4.1cm}
\begin{center}
Fig.7: Second order approximation to the Wigner quantum \\
trajectories: numerical solution of eq.(85) with $\epsilon=0.25$
and $\hbar=m=1$.
\end{center}

\end{document}